\documentclass[usenatbib]{mn2e}
\usepackage{graphicx}
\usepackage{times}
\usepackage{color}

\title{The angular momentum of baryons and dark matter halos revisited}
\author[T. Kimm et al.]{
\parbox[t]{\textwidth}{
Taysun Kimm$^{1}$\thanks{e-mail: taysun.kimm@astro.ox.ac.uk}, Julien
Devriendt$^{1,2}$, Adrianne Slyz$^1$, Christophe Pichon$^{3,1,2}$, Susan A.
Kassin$^4$\thanks{NASA Postdoctoral Program Fellow}, and Yohan Dubois$^1$
}
\vspace*{6pt} \\
$^1$ Department of Physics, Denys Wilkinson Building, Keble Road, Oxford, OX1 3RH, United Kingdom\\
$^2$ CRAL, Universit\'e Claude Bernard Lyon I, CNRS UMR 5574, ENS-Lyon, 9 Avenue Charles Andr\'e,
F-69561 Saint-Genis Laval, France\\
$^3$ Institut d'Astrophysique de Paris, Universit\'e Pierre et Marie Curie Paris 6, CNRS, UMR 7095, F-75014, Paris, France \\
$^4$ Astrophysics Science Division, Goddard Space Flight Center, Code 665, Greenbelt, MD 20771, USA\\}
\begin{document}
\maketitle

\def\lesssim{\lower.5ex\hbox{$\; \buildrel < \over \sim \;$}}
\def\gtrsim{\lower.5ex\hbox{$\; \buildrel > \over \sim \;$}}
\newcommand{\msun}{\mbox{$M_\odot$}}
\newcommand{\mvir}{\mbox{$M_{\rm vir}$}}
\newcommand{\rvir}{\mbox{$R_{\rm vir}$}}
\newcommand{\mn}{\mbox{{\sc \small Horizon}-MareNostrum}}
\newcommand{\nut}{\mbox{{\sc \small Nut}}}
\newcommand{\ramses}{\mbox{{\sc \small Ramses}}}
\newcommand{\jtot}{\mbox{$j$}}
\newcommand{\jz}{\mbox{$j_{\rm z}$}}
\newcommand{\Jtot}{\mbox{$J$}}
\newcommand{\tlb}{\mbox{$t_{\rm lb}$}}
\newcommand{\jgas}{\mbox{$j_{\rm gas}$}}
\newcommand{\jgasout}{\mbox{$j_{\rm gas,out}$}}
\newcommand{\jgascen}{\mbox{$j_{\rm gas,cen}$}}
\newcommand{\jgasoutv}{\mbox{${{\vec\mathbf j}}_{\rm gas,out}$}}
\newcommand{\jgascenv}{\mbox{${{\vec\mathbf j}}_{\rm gas,cen}$}}
\newcommand{\jdm}{\mbox{$j_{\rm dm}$}}
\newcommand{\jstar}{\mbox{$j_{\rm star}$}}
\newcommand{\jbar}{\mbox{$j_{\rm bar}$}}
\newcommand{\blue}{\color{blue}}

\newcommand{\aap}{\mbox{A\&A}}
\newcommand{\apj}{\mbox{ApJ}}
\newcommand{\apjl}{\mbox{ApJL}}
\newcommand{\apjs}{\mbox{ApJS}}
\newcommand{\mnras}{\mbox{MNRAS}}
\newcommand{\nat}{\mbox{Nature}}

\begin{abstract}
Recent theoretical studies have shown that galaxies at high redshift are fed by cold, 
dense gas filaments, suggesting angular momentum transport by gas differs from that 
by dark matter. Revisiting this issue using high-resolution cosmological hydrodynamics 
simulations with adaptive mesh refinement (AMR), we find that at the time of accretion, 
gas and dark matter do carry a similar amount of specific angular momentum, 
but that it is systematically higher than that of the dark matter halo as a whole.
At high redshift, freshly accreted gas rapidly streams into the central region of the halo, 
directly depositing this large amount of angular momentum within a sphere of radius $r = 0.1 \rvir$.
In contrast,  dark matter particles pass through the central region unscathed, 
and a fraction of them ends up populating the outer regions of the halo ( $r/\rvir >0.1$), 
redistributing angular momentum in the process. As a result, large-scale motions of the 
cosmic web have to be considered as the origin of gas angular momentum rather 
than its virialised dark matter halo host. This generic result holds for halos of all masses 
at all redshifts, as radiative cooling ensures that a significant fraction of baryons remain trapped 
at the centre of the halos. Despite this injection of angular momentum enriched gas,  
we predict an amount for stellar discs which is in fair agreement with observations at z=0. 
This arises because the total specific angular momentum of the baryons (gas and stars) 
remains close to that of dark matter halos. Indeed, our simulations indicate that any 
differential loss of angular momentum amplitude between the two components is minor
even though dark matter halos continuously lose between half and two-thirds of their 
specific angular momentum {\it modulus} as they evolve. In light of our results, a substantial revision 
of the standard theory of disc formation seems to be required. We propose a new scenario 
where gas efficiently carries the angular momentum generated by large-scale structure 
motions deep inside dark matter halos, redistributing it only in the vicinity of the disc.
\end{abstract}


\voffset=-0.6in
\hoffset=0.2in

\section{Introduction}
Along with mass, angular momentum growth shapes the fundamental properties of galaxies.
Using linear theory, one can show that density perturbations acquire
angular momentum through their asymmetric interactions with the larger
scale tidal field \citep{peebles69,doroshkevich70,white84}. 
Thus one expects the angular momentum of the Lagrangian region encompassing a future 
virialized structure to scale like $a^2(t) \dot{D}(t)$ until it decouples from the Hubble expansion, 
where $a$ and $\dot{D}$ are the scale factor of the universe and the linear growth rate of
density perturbations respectively \citep{white84}. Further assuming that the gas experiences 
the same tidal field as the dark matter, they should initially share their specific angular momentum. 
Provided this specific angular momentum is conserved as the gas radiatively cools,
early studies were able to reasonably match crucial observed relations, such as the 
Tully-Fisher relation or the size-rotation velocity relation \citep{fall80,dalcanton97,mo98}.
In spite of several severe shortcomings in the theory pointed out by authors like \cite{hoffman86} 
(including no account of secondary infall and mergers between virialized objects), 
this success encouraged all (semi-analytic) galaxy formation models published
to date \citep[e.g.][]{cole00,hatton03,croton06,monaco07,somerville08,dutton09,khochfar11}
to rely on the core assumption that gas and dark matter contained within
the same virialized structure split specific angular momentum equally.   
However, the recent findings that the vast majority of galaxies are mainly fed gas 
by cold, thin and dense filaments which penetrate deep inside the virial radius of
their host dark matter halo \citep[e.g.][]{keres05,ocvirk08,dekel09,powell11}
begs for a reassessment of the validity of the classical angular momentum scenario.

According to the standard picture of galaxy formation, gas is first shock-heated to the virial 
temperature of its dark matter halo host as this latter collapses, and subsequent accreted material
encounters the resulting virial shock as it penetrates the halo \citep{rees77,silk77,white78}.  
Since the shock-heated gas is more or less spatially re-distributed as the dark matter before it can 
cool, one is led to logically postulate that its specific angular momentum closely tracks that of the 
dark matter halo. Non-radiative hydrodynamics cosmological simulations have confirmed that this 
is the case \citep[e.g.][]{van-den-bosch02}.

However, as soon as the gas is permitted to radiatively cool and form stars, this tight correlation 
between gas and dark matter spin should break down. For instance, a fraction of gas accreted at 
earlier times with lower angular momentum will be converted into stars, elevating the specific 
angular momentum of the remaining gas \citep{dutton09}. Meanwhile the specific angular momentum of the dark
matter halo is expected to either be frozen in after turn around or decline due to the presence of 
external torques \citep[c.f.][]{peebles69,book11}. Hence it is not very surprising that radiative 
hydrodynamics cosmological simulations show a difference in the spin parameters between gas 
and dark matter. Indeed, \citet{stewart11} have recently reported that for a pair of resimulated 
halos, cold gas had 3 to 5 times more specific angular momentum than dark matter  
\citep[see also][]{sales10}. In this context, the questions which naturally arise are: (i) what are the 
mechanisms that segregate dark matter and gas angular momenta? (ii) is this segregation 
universal or does it depend on halo properties?  and (iii) does it evolve with redshift? Since the 
gas is the main agent for transferring angular momentum from (super) halo scales down to central 
galaxies, it is key to investigate the evolution of its angular momentum in the region extending 
between the galaxy and the virial radius of the dark matter halo (i.e. $0.1\rvir  < r \leq \rvir$). 

Within galaxy size virialised halos, the picture of gas accretion has recently been significantly 
revised, with the rediscovery of an idea first put forward by \citet{binney77} that infalling gas is 
never shock-heated to the virial temperature but instead flows through an `isothermal' shock, 
reaching the galactic disc cold \citep{katz03,birnboim03,keres05,ocvirk08,brooks09}. Moreover, 
this gas is brought along anisotropic narrow streams which persist deep within the halo, contrary 
to their broader dark matter counterparts which rapidly fade away at the virial radius 
\citep{powell11}. Therefore, one wonders how differently this filamentary gas advects angular 
momentum throughout the halo. 

With such questions in mind, we use a set of high-resolution cosmological simulations to revisit  
two key assumptions of disc formation theory, namely that gas carries the same amount of 
specific angular momentum as its host dark matter halo and that this specific angular momentum 
is conserved as this material gets accreted onto the central disc. The details of the simulations 
are described in Section 2. In Section 3, we present the time evolution of angular momentum of 
baryons and non-baryonic matter along with a comparison to observational data at $z=0$. 
We also discuss in this section how our results depend on halo properties and evolve with 
redshift. Finally, we discuss the discrepancy between dark matter and gas specific angular 
momenta in Section 4 and conclude in Section 5.

\begin{figure}
\centering
   \includegraphics[width=7.5cm]{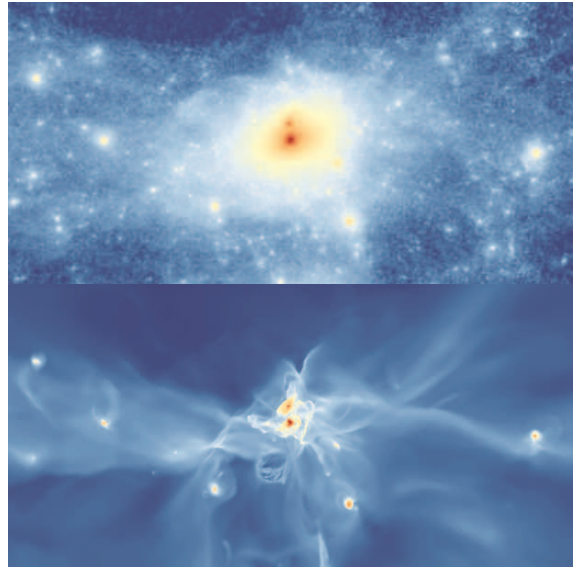}
   \caption{Projected densities of the dark matter (top) and gas (bottom) centred on a halo of 
   $\simeq10^{11}\msun$ at $z=3$ from the \nut\ simulation with supernova feedback (NutFB). 
   A $6\rvir$ region is projected and a figure with aspect ratio 1:2 is displayed, where the longer 
   length corresponds to $=6\rvir$ ($=$ 224 kpc). At $z=3$ the scale-height of the disc as well as 
   the large scale filamentary structure is well resolved in the simulation. It can be seen that dark 
   matter filaments are broader than gas filaments. The gas filaments are not destroyed by 
   supernova explosions.}
   \label{fig:snapshot}
\end{figure}

\section{Simulations}

We use the Eulerian AMR code, \ramses\ \citep{teyssier02}, to investigate the angular momentum 
evolution of baryons and dark matter. \ramses\  uses a second-order Godunov scheme to solve 
the Euler equations, and an adaptive particle-mesh method to solve the Poisson equation.
Since the outcome of hydrodynamics simulations is subject to resolution and physical ingredients,
we make use of five simulations to draw robust conclusions. These simulations include the \nut\ series (\citealt{powell11}), the \mn\ simulation \citep{ocvirk08,devriendt10}, and two other 
cosmological simulations (Cosmo25, Cosmo50, Dubois et al. {\sl in prep.}) of fairly large volumes 
but with lower resolution than the \mn\ simulation. Whereas simulations in the \nut\ series focus on 
the evolution of an individual, Milky Way-like galaxy, the \mn, Cosmo25 and Cosmo50 simulations 
give the statistical properties of galaxies spanning a wide range of halo mass. Physical ingredients 
common to most of our simulations include reionisation, cooling, star formation and supernova 
feedback. These are summarised in Table 1, along with simulation parameters.

To model the reionisation of the Universe, a uniform UV background radiation field  (modelled as 
a heating term in the energy equation) is turned on at high redshift 
(see Table~\ref{SimSummary}) following \citet{haardt96}. Gas dissipates energy through atomic 
cooling down to 10$^4$K \citep{sutherland93}. For the \nut\ simulations metal line cooling can 
lower the gas temperature further (below $\sim 10$ K). When the gas density in a grid cell 
exceeds a given threshold ($n_{\rm th}$, see Table 1), star particles are spawned by a Poisson 
process according to a Schmidt law with a 1\% efficiency of the star formation per free-fall time 
\citep{kennicutt98,dubois08}. This threshold density is chosen so that it is inferior or equal to the 
maximal Jeans density reached on the finest level.  In the simulations with supernova feedback, 
after $\sim 10$ Myrs, massive stars undergo Type II supernova explosions, releasing half of their 
$10^{51}$ ergs into their surroundings as kinetic energy and the other half as thermal energy 
\citep{dubois08}. During this phase, processed heavy elements are dispersed, enriching the 
interstellar and intergalactic medium. In what follows, we elaborate on the details of each 
simulation.

\begin{table*} 
\caption{Summary of simulation parameters and physical ingredients. From left to right, columns 
are as follows: simulation name, comoving box size, number of dark matter particles, level of the 
root grid, level of the finest grid, minimum grid size, dark matter particle mass, star particle mass, 
threshold density for star formation, redshift down to which the simulation is carried out, indication 
of whether supernova feedback is included, redshift at which UV background heating is initiated, 
and remarks.}
\label{SimSummary} 
\begin{tabular}{@{}lcccccccccccc}
\hline 
Simulations &    L     & $N_{\rm DM}$ &  $l_{\rm min}$ & $l_{\rm max}$ & $\Delta x_{\rm min}$&  $m_{\rm DM}$ & $m_{\rm star}$ & $n_{\rm th}$  & $z_{\rm end}$ & SN & UV & Remarks\\ 
                   & (Mpc/h) &                           &                          &                           &      (pc)               & (\msun)               & (\msun)              &  (${\rm H/cm^{3}}$) &       &      &        & \\ 
\hline 
NutFB & 9 & $1024^3$ & 7 & 20 & 12 & $5\times 10^4$ & $2\times 10^4$ & 400 & 3 & Y & z=8.5 & Resimulation\\
NutCO & 9 & $1024^3$ & 7 & 18 & 48 & $5\times 10^4$ & $3\times 10^4$ & 10 & 0 & -- & z=8.5 & Resimulation\\
NutAD & 9 & $1024^3$ & 7 & 18 & 48 & $5\times 10^4$ &           --            &  --   & 0 &  --  & z=8.5 & Resimulation\\
MareNostrum & 50 & $1024^3$ & 10 & 16 & 1090 & $1\times 10^7$ & $2\times 10^6$ & 0.1 & 1.5 & Y & z=8.5 & \\
Cosmo25 & 25 & $256^3$ & 8 & 15 & 1090 & $8\times 10^7$ & $4\times 10^6$ & 0.1 & 0 & Y & z=10.5 & \\
Cosmo50 & 50 & $256^3$ & 8 & 15 & 2180 & $6\times 10^8$ & $3\times 10^7$ & 0.1 & 0 & Y & z=10.5 & \\
\hline 
\end{tabular} 
\end{table*}

The \nut\ series make use of the resimulation (also called `zoom') technique to follow the evolution 
of a Milky Way-type galaxy in a $\Lambda$CDM cosmology. \citet{powell11} reported on high 
redshift results (z $\geq$ 9) from the ultra-high resolution resimulations in the \nut\ series. These 
reached a maximum physical spatial resolution of $\sim$ 0.5 pc at all times. To track the evolution 
of the galaxy down to lower redshifts, the \nut\ suite also includes three resimulations with lower 
{\em spatial} resolution (but identical DM particle mass resolution) and the following physics: (i) 
adiabatic with a uniform UV background turned on instantaneously at z$=8.5$ (NutAD) (ii) cooling, 
star formation, and UV background (NutCO) and (iii) same as (ii) but with supernova feedback and 
metal enrichment (NutFB). NutAD and NutCO have maximum 48 pc (physical) resolution at all 
times and reach $z=0$, whereas NutFB has maximum 12 pc (physical) resolution and only 
reaches $z=3$. Fig.~\ref{fig:snapshot} shows a snapshot of the DM (upper panel) and gas 
(lower panel) of a region centred on a halo hosting a Milky Way-type galaxy at $z=3$ in NutFB.

We recall the important details of the \nut\ resimulations here. 
The simulation volume is a 9$h^{-1}$ Mpc comoving periodic box evolving according to a 
WMAP5 cosmology \citep{dunkley09} ($\sigma_8=0.8$, $\Omega_m=0.258$, $\Omega_{\Lambda}=0.742$,
$h\equiv  H_0/(100 {\rm km s^{-1} Mpc^{-1}}) = 0.72$). Initial conditions are generated using 
{\texttt{MPgrafic}} \citep{prunet08}, a parallel version of the {\texttt Grafic} package 
\citep{bertschinger01}. Within this volume we identify the region where a Milky Way-like
galaxy (halo virial mass of $\simeq$ 5 $\times$10$^{11}$M$_{\odot}$ at $z=0$) will form. 
This region encompasses a volume of side length $\sim$ 2.7 h$^{-1}$ Mpc. While the root grid 
for the entire simulation volume is 128$^3$, within the ($\sim$2.7 h$^{-1}$ Mpc)$^3$ region, 
we place an additional three nested grids, giving an equivalent resolution of 1024$^{3}$ dark 
matter particles, each with mass M$_{\rm DM}$ $\simeq 5\times 10^4 \msun$. To fix the 
maximum physical resolution to a constant value (12 pc for NutFB and 48 pc for NutCO and 
NutAD) as the universe expands and the simulation evolves, we further refine on the finest 
fixed grid within the 2.7 h$^{-1}$ Mpc$^3$ region according to a quasi-Lagrangian strategy, 
i.e. when the number of dark matter particles in a cell reaches 8 or equivalently when the 
baryon plus dark matter density in a cell increases by a factor of 8. 
Table~\ref{SimSummary} lists the maximum level triggered for each simulation. 
Because of the higher spatial resolution in NutFB it uses a higher density threshold for 
star formation ($n_{\rm H,th} = 400 ~{\rm cm^{-3}}$) than the NutCO run 
($n_{\rm H,th} = 10 ~{\rm cm^{-3}}$). The star particle mass ($\simeq 2-3 \times10^4\msun$) is 
determined by the combination of minimum grid size and density threshold for star formation 
\citep{dubois08}. In NutFB, we assume that every supernova bubble with an initial radius
of 32 pc sweeps up the same amount of gas as that initially locked in the star particles. This is 
usually expressed as a mass loading factor of unity ($\eta=1$).

The other three simulations are large volume cosmological simulations, and as such are performed 
with lower spatial (1--2 kpc) and mass resolution. More specifically, the mass of each dark matter 
particle is $m_{\rm DM}\simeq 10^7\msun$ for the \mn\ simulation, $8\times 10^7\msun$ for the 
intermediate size run (Cosmo25) and $6\times 10^8\msun$ for the Cosmo50 run. As in the NutFB 
run, supernova feedback and UV background heating are included in the simulations, but the 
radius of the initial supernova bubble is set to twice the minimum size of the grid (see 
Table~\ref{SimSummary}). Note also that the adopted cosmology for the \mn\ simulation (WMAP1) 
is different from the others, but as we will show this has very little impact on our results, if at all. 
We refer interested readers to \citet{ocvirk08} and \citet{devriendt10} for a detailed description 
of the \mn\ simulation set-up. 

In all the simulations, we identify (sub) haloes using the {\sc \small AdaptaHop} algorithm 
\citep{aubert04}, which is based on the detection of peaks and saddle points in the dark matter 
density field, supplemented by the most-massive subhalo algorithm developed by \citet{tweed09}. 
The virial radius of halos is defined as the maximal radius within which the virial theorem is 
satisfied to better than 20\%. We further define gas belonging to a satellite galaxy as gas residing 
within the half-mass radius of its host DM satellite halo. The centre of a halo, which we use to 
compute angular momentum, is defined as the centre of mass of dark matter and baryons. The 
mean motion of the halos is determined by computing the centre of mass velocity of dark matter 
particles, gas, and stars within their virial radii.

 \section{The angular momentum of virialized halos}

To understand how galaxies acquire their angular momentum, we study the angular momentum 
evolution of the different components (gas, dark matter, stars) inside their host halos. We compute 
the specific angular momentum vectors as: 
\begin{equation}
{\vec \mathbf j} =  \frac{\sum_i m_i \vec{\mathbf{r}}_i \times \vec{\mathbf{v}}_i}{\sum_i m_i}, 
\label{eq:defj}
\end{equation}
where $\vec{\mathbf{r}}_i$ is the radial distance from the centre of mass of the halo (includes dark 
matter and baryons),  $\vec{\mathbf{v}}_i$ is the peculiar velocity and $m_i$ is the mass of the 
i-th dark matter (star) particle or gas cell. In what follows we use different subscripts to denote the 
specific angular momentum of different components in different regions of the halo. These are 
summarised in Table 2.

\begin{table}
   \centering
   \caption{Notations (first column) for specific angular momenta used in this study. The first 
   subscript indicates the component used for the calculation (DM, gas, stars, baryons). The 
   second column specifies the region (normalised to the halo virial radius) over which the 
   specific angular momentum is calculated. The default region over which we measure specific 
   angular momentum is a sphere extending from the centre of mass of the halo to the virial radius. 
   A second subscript indicates whether we exclude ($\jgasout$) or include ($\jgascen$) the gas in 
   the central region ($r/\rvir \leq 0.1$) of the halo for the specific angular momentum calculation. 
   The third column lists the components included in the specific angular momentum calculation. }
   \begin{tabular}{@{} ccc} 
   \hline
   Notation & Spatial extent & Component \\
   \hline
      \jdm & $ r/\rvir \leq 1$ & dark matter\\
         \jgas & $  r/\rvir \leq 1$ & gas\\
   \jgasout & $0.1 \leq r/\rvir \leq 1$ & gas\\
   \jgascen & $r/\rvir \leq 0.1$ & gas\\
   \jstar & $ r/\rvir \leq 1$ & all stars except those in satellite galaxies\\
      \jbar & $ r/\rvir \leq 1$ & all stars (satellite galaxies included) + gas\\
   \hline
   \end{tabular}
   \label{tab:booktabs}
\end{table}

\begin{figure}
   \centering
 \includegraphics[width=8.5cm]{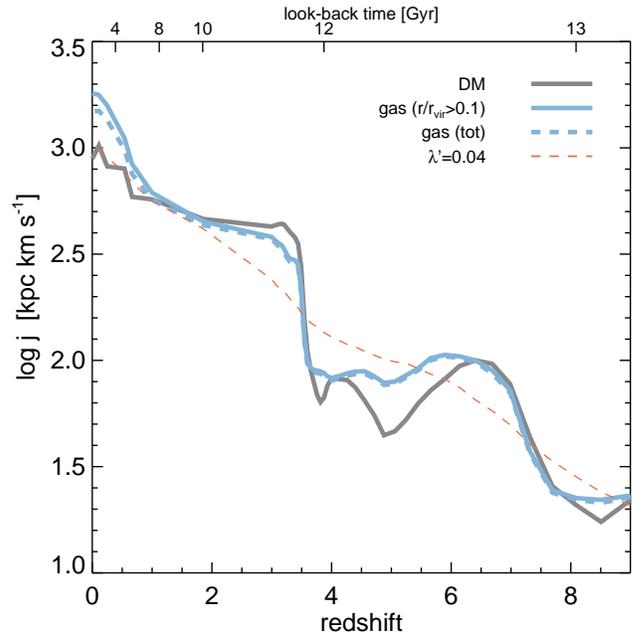} 
   \caption{Evolution of the specific angular momentum (\jtot) of a dark matter halo hosting a 
   Milky Way-type galaxy in the adiabatic run (NutAD). Different colours denote different 
   components: dark matter (\jdm, thick grey), halo gas (\jgasout,   $0.1\leq r/\rvir \leq 1$, thick blue).  
   Also plotted as a blue dashed line is \jtot\ of all the gas inside the virial radius, \jgas. We also 
   display the specific angular momentum corresponding to a reduced spin value 
   $\lambda'=0.04$ (red dashed line) for the dark matter halo. The specific angular momentum of 
   the halo gas (\jgasout) closely follows that of the dark matter halo (\jdm) in this non-radiative 
   simulation. }
   \label{fig:jtot_ad}
\end{figure}

\subsection{Adiabaticity and the cosmic origin of angular momentum} 

We begin our investigation of the evolution of angular momentum of the various components of a 
virialized halo with the simplest (in terms of physics) high resolution NutAD run. Since gas cannot 
cool radiatively in this simulation, newly accreted material is shock heated by the pressure 
supported intra-halo medium. As a result, its radially oriented initial velocity is isotropized, and 
drives the gas density field towards spherical symmetry.  Accreted dark matter particles are also 
more or less isotropically redistributed within the host halo by the collisionless violent relaxation 
process. Given that (i) gas and dark matter within the halo experience the same larger-scale 
torques \citep[e.g.][]{peebles69,book11}, and (ii) gas and dark matter are driven by the gravitational 
collapse to a very similar equilibrium distribution (i.e. to a good approximation that of an isothermal 
sphere since the total amount of angular momentum provided by tidal torques is very limited), we 
expect them to have similar \jtot. Indeed, Fig.~\ref{fig:jtot_ad} shows that \jgasout\ closely tracks 
\jdm\ regardless of whether or not mergers, easily identified by large jumps in \jtot, occur. When 
the dark matter angular momentum is expressed in terms of the spin parameter  
($\lambda'=\jdm/\sqrt{2}\rvir V_{\rm c}$, \citealt{bullock01}), we recover the typical value of 
$\lambda'\simeq 0.04$ {\em at all times} with fluctuations of up to a factor of $\sim 2$ around this 
value.

An interesting feature in Fig.~\ref{fig:jtot_ad} is that \jdm\ increases with time (albeit at a reduced 
rate below $z=3$), implying that late infall carries a larger amount of angular momentum. This is 
not a completely unexpected result given that dark matter haloes are known to experience little 
evolution of spin parameter with time \citep[e.g.][]{peirani04}. As their mass and radius grow, so 
must their angular momentum. However, it is not trivial to understand {\em why} the late accretion 
has larger \jtot. A naive answer is that material with larger angular momentum takes more time to 
reach the potential well, but this does not explain {\em how} it acquired such a large angular 
momentum in the first place. We discuss in detail the cosmic evolution of the angular momentum of 
halos (which necessarily goes beyond tidal torque theory)  in a companion paper \citep{pichon11}, 
but for completeness' sake, we briefly outline the main idea in the paragraph below.

\begin{figure*}
   \centering
   \includegraphics[width=8.5cm]{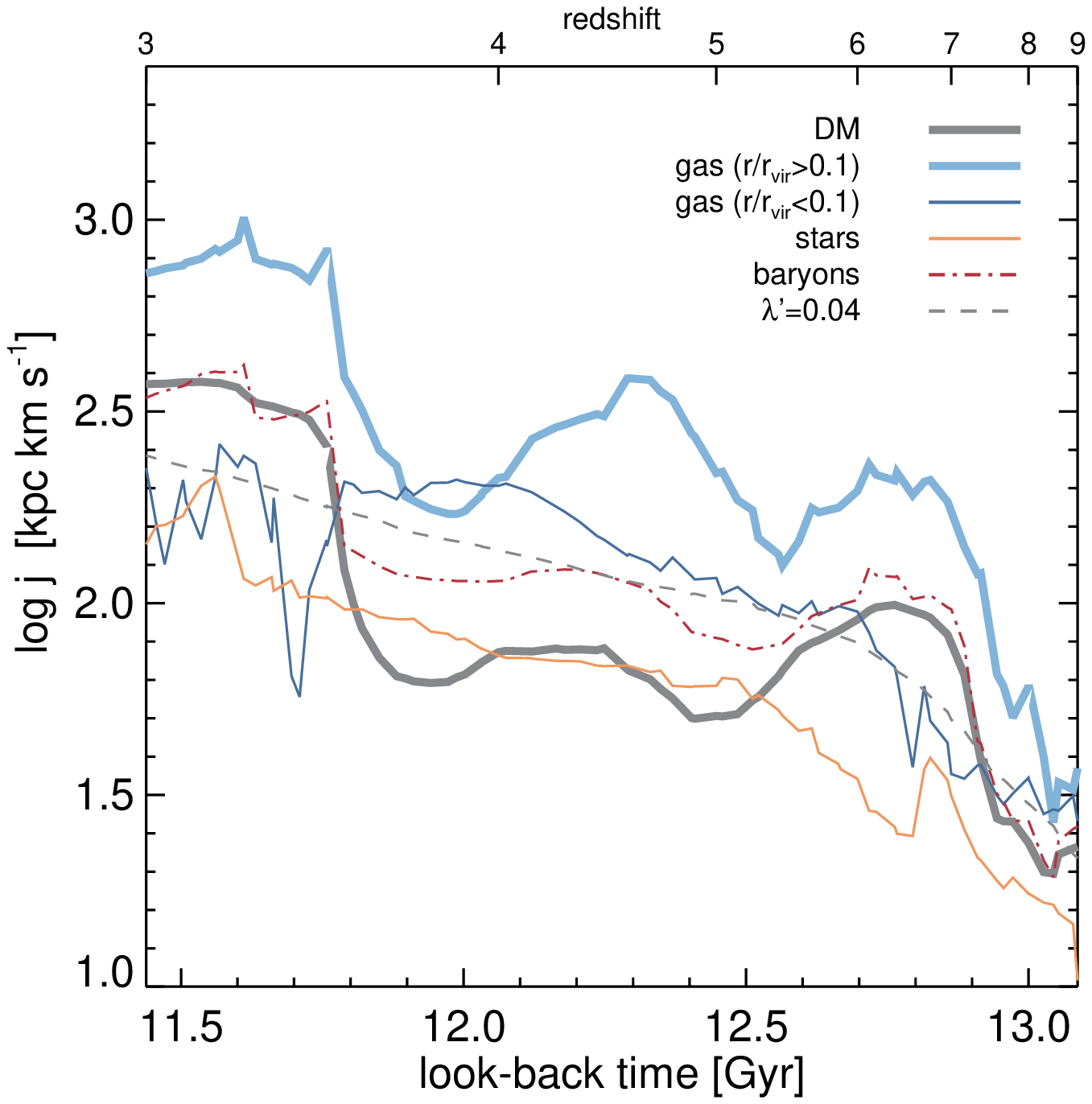} 
      \includegraphics[width=8.5cm]{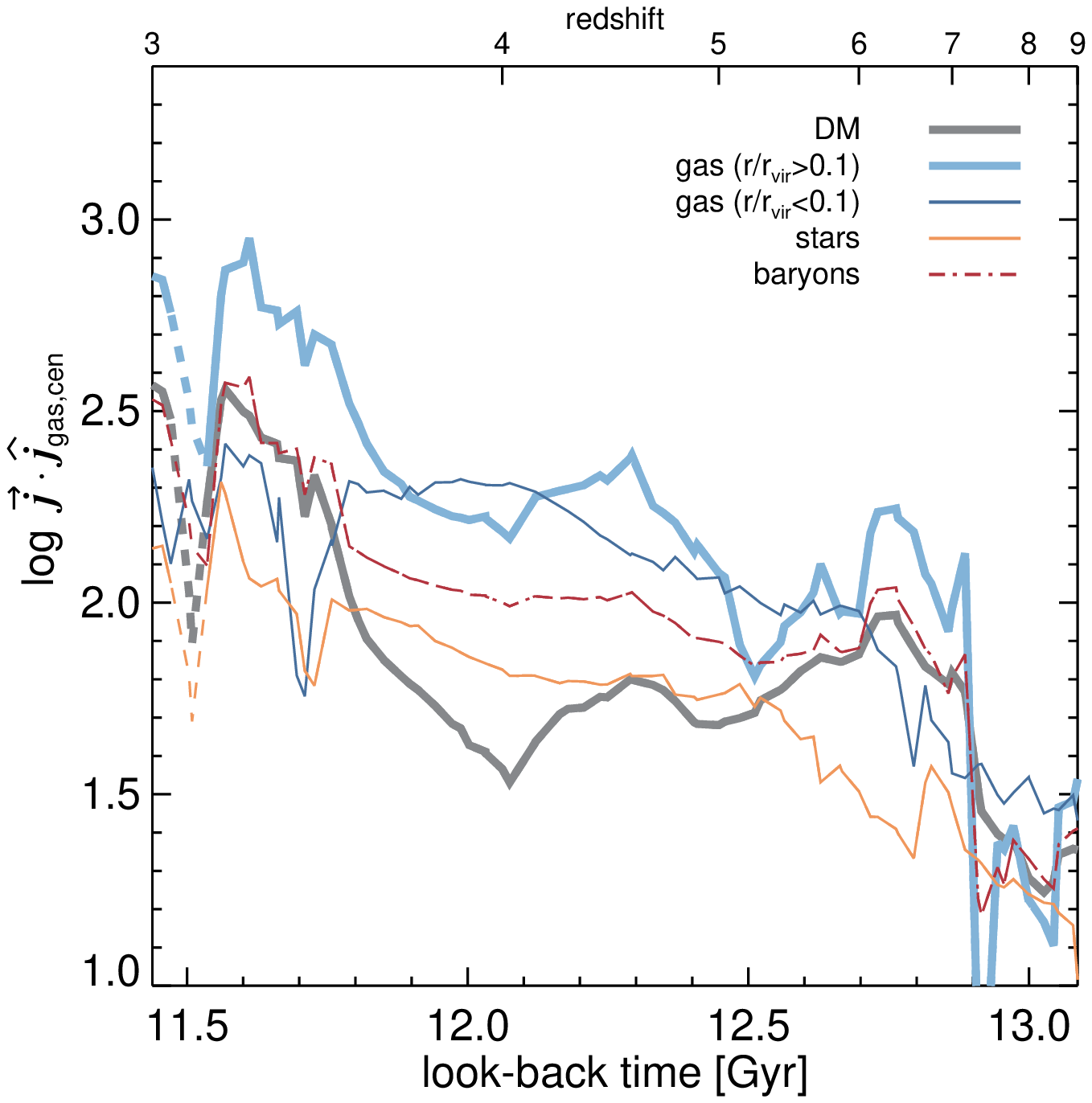} 
   \caption{Left panel: evolution of the specific angular momentum modulus (\jtot) from the NutFB 
   run. Different colours denote different components: dark matter (\jdm, thick grey), gas at 
   $0.1\leq r/\rvir \leq 1$ (\jgasout, thick blue), central gas ($r/\rvir<0.1$, \jgascen, thin blue), 
   stars (\jstar, thin orange), all baryons in the halo (\jbar, dot-dash red). Note that the specific 
   angular momentum calculations for dark matter (\jdm), gas (\jgasout) and baryons (\jbar) 
   includes the contribution from satellite galaxies, while it is excluded for the measurement 
   of \jstar\ and \jgascen. Angular momentum corresponding to a constant reduced spin value 
   of $\lambda'=0.04$ is included as a grey dashed line. The specific angular momentum of gas 
   (\jgasout) is larger than that of its host dark matter halo (\jdm), suggesting that the angular 
   momentum of the gas is not acquired during the collapse of this latter. Right panel: specific 
   angular momentum along the spin axis of the gas in the central region ($\hat{j}_{\rm gas,cen}$). 
   By definition, $\vec{j}\cdot\hat{j}_{\rm gas,cen}$ for gas in the central region in this panel (thin 
   blue line) is the same as \jgascen\ in the left panel.  We denote negative values as dashed lines. 
   Different angular momentum amplitudes between lines of the same type in the left and right 
   panels indicate the misalignment between the central gas and each component.
   }
   \label{fig:jtot}
\end{figure*}

The dynamics of the gas and dark matter flowing along what has been dubbed the `cosmic web' 
can be understood as the anisotropic time evolution of the initial gaussian random gravitational 
field. Cosmological structures originate as peaks in the associated initial density field 
\citep{bond96}, which are connected to other peaks through peak patches. The bulk motion of a 
peak patch is determined by the gradient of the large-scale potential. The latter also drives the 
motions of filaments which exist at the intersection of at least three void patches and connect 
peaks. As a consequence of the asymmetry between voids, the gas and dark matter flowing out of 
these voids acquire a transverse velocity when they intersect at a filament. This transverse velocity 
is the seed of a halo's angular momentum which is then advected along the filaments all the way 
into the halo sitting on the peak. Since the transverse velocity along a filament is constant to zeroth 
order approximation, the material initially located further away from the peak will naturally 
contribute more angular momentum. Note that the infall of matter along such filaments will 
{\em coherently} contribute an increasing amount of angular momentum as time goes on 
(shown in Fig.~\ref{fig:jtot_ad}), since the filament preserves its orientation over long stretches 
of time.

\subsection{Radiative cooling, supernovae feedback, and the evolution of angular momentum at high redshift} 
\label{sec:highz}

In the real Universe, gas can radiatively cool and form stars, which should therefore be considered
as a sink for gas mass and angular momentum. Moreover, at high redshift, star formation is very 
active \citep{bouwens09} and the effect of supernova explosions on the evolution of the halo gas 
may prove important. Thus, we make use of the high-resolution NutFB run which includes these 
processes to investigate the evolution of angular momentum of a DM halo at high redshift 
($z \geq 3$).

The left panel of Fig.~\ref{fig:jtot} shows that \jgasout\ and \jdm\ increase steadily, although with 
occasional rapid variations located around look-back times ($\tlb$) of $\tlb=$12.9, 12.4, and 11.7 
Gyr. These angular momentum `jumps' are generated by episodes of enhanced accretion of dense 
gas contained in satellite galaxies and filaments \citep[e.g.][]{maller02,peirani04}. 
Indeed, the most significant {\em halo} merger 
occurs at $\tlb \simeq 11.7$ Gyr (satellite to host halo mass ratio $M_{\rm sat}/M_{\rm host}\sim 
0.1$) and the majority of the other 12 minor mergers ($0.01 \le M_{\rm sat}/M_{\rm host}\le 0.1$) 
take place between $12.5 \lesssim \tlb \lesssim 13.0$ Gyr. In addition to these mergers, material 
accreted along filaments also contributes to the rapid variation of \jtot\ since the density in the gas 
filaments varies (see also Fig.~2 of \citealt{brooks09}). 
 
However, the key feature of Fig.~\ref{fig:jtot} (left panel) is that \jgasout, although tightly correlated 
with \jdm, is always larger by a factor of a few \citep[see also][]{chen03,sharma05,gottlober07,sales10,stewart11}. 
This is in contradiction with the picture in which the angular momentum of the halo gas is acquired 
during the virialisation of its host dark matter halo and suggests that if the gas and the DM halo 
interact at all and exchange angular momentum during the collapse, the effect is likely to be 
unimportant, at least in the outer regions of the halo. Indeed this is more in line with the picture 
according to which gas accreted at the virial radius supersonically streams (almost) directly into the 
central galaxy \citep[e.g.][]{brooks09,powell11}. Moreover,  Fig.~\ref{fig:jtot} (left panel) also shows 
that \jgascen\ (thin blue solid line) is systematically smaller than \jgasout\ (thick blue solid line).  
There are two possible explanations for this. First, the specific angular momentum {\em modulus} 
of the gas is not conserved during its gravitational collapse and subsequent accretion onto the 
central galaxy because \jgasoutv\ is misaligned with \jgascenv\ \citep[also see][]{van-den-bosch02,hahn10,bett10}. 
The second possibility is that unlike the gas in the outer part of the halo which yields an 
instantaneous picture of the accreted angular momentum, the central gas reflects the angular 
momentum accretion {\em history} of the gas. We now discuss each of these possibilities in turn.

The difference in angular momentum alignment between outer and inner gas is indicated in the 
right panel of Fig.~\ref{fig:jtot} where we plot the component of \jgasoutv\ (thick solid blue line) parallel to 
\jgascenv. This figure clearly shows that this component is remarkably well preserved (as one would expect 
for an isolated axisymmetric system) and it is therefore tempting to conclude that misalignment accounts for 
most of the difference between \jgasout\ and \jgascen. However, the time evolution of the projected 
component of  \jgasoutv\ along \jgascenv\ direction does not correlate very well with the time evolution of 
\jgascen.  This behaviour is naturally interpreted as a gas accretion history effect. Indeed, as the mass of gas 
in the low amplitude specific angular momentum central region ($r/\rvir \leq 0.1$)  is comparable 
(between a factor 2-3 smaller) to that in the outer reservoir ($0.1 \leq r/\rvir \leq 1$)  (see Appendix 
for details on the evolution of the gas mass in these regions), even if \jgasoutv\ was perfectly aligned 
with \jgascenv, the resulting modulus of the specific angular momentum of the inner region after 
accretion would naturally be reduced compared to that of the freshly accreted gas. As a result, 
notwithstanding mergers, \jgascen\ grows more smoothly than \jgasout\ and takes more time to 
reach a given level of specific angular momentum. This integrated effect is exacerbated for the 
stellar component, which displays an even smaller amount of specific angular momentum than 
\jgascen. Since the global star formation timescale is much longer than the dynamical timescale of 
the central region  \citep[e.g.,][]{kennicutt98}, most of the stars will form from older gas, with a lower 
angular momentum amplitude. Even though newly formed stars carry more angular momentum, 
they represent a smaller fraction of the total stellar mass  than the central gas mass compared to 
that of the total halo gas.  Therefore, a supplementary time delay to reach a given level of \jtot\ and 
a smoother growth in specific angular momentum amplitude of the stellar component is inevitable 
\citep{dutton09}.

Finally, when we measure the specific angular momentum of the {\em total} baryonic component, 
i.e. gas and stars, (\jbar: thin red dot-dashed line in the left panel of Fig.~\ref{fig:jtot}), we find that its 
amplitude most closely tracks that of the specific angular momentum of the DM (\jdm, thick grey 
solid line) which we know reflects the mass assembly history of the halo. We are not implying that 
$|{\vec \mathbf j}_{\rm dm}|$ of individual dark matter particles is conserved during the collapse, 
but simply that, as the baryon angular momentum is advected deep within the halo, we can only 
probe the accretion history of 
the gas by including all baryons within the inner region. When this effect is accounted for, there is no 
indication in our simulation that baryons lose more (or less) specific angular momentum amplitude than DM 
during gravitational collapse of the halo. Interestingly,  the specific angular momentum of the central 
gas has a value close to that of dark matter for virialised halos typically measured in N-body simulations 
($\lambda' = 0.04$) \citep[e.g.][]{efstathiou79,barnes87}. It should be noted, however, that it is 
established with the aid of gas within $0.1 \leq r/\rvir \leq 1$, which has a larger \jtot\ than the dark matter. 
We therefore surmise that if this gas had started with the same specific angular momentum 
as the dark matter halo to begin with, the central gas would have been left with roughly half of the 
typical DM specific angular momentum. We come back to the crucial issue of specific angular 
momentum segregation between gas and DM in Section \ref{sec:diff} but now address the question 
of the importance of feedback in regulating \jgascen.

\begin{figure}
   \centering
   \includegraphics[width=8.5cm]{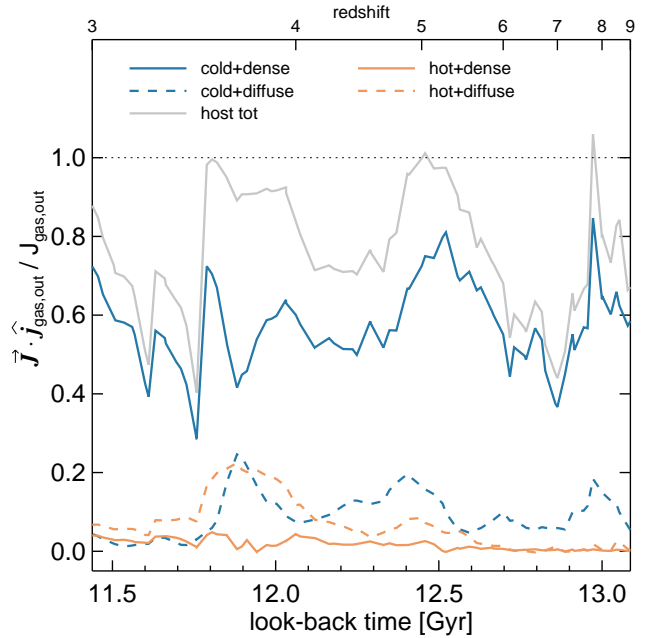} 
   \caption{Fraction of total angular momentum of gas at $0.1\leq r/\rvir \leq 1$ 
   from various components. We divide the gas into four different phases: i) cold ($T<10^5$K), 
   dense filamentary ($n_{\rm H, fil}\leq n_{\rm H} < n_{\rm H,th}$), ii) cold diffuse 
   ($n_{\rm H} < n_{\rm H, fil}$), iii) hot ($T>10^5$K) dense, and iv) hot diffuse component (see 
   Equation~(\ref{eq:deffil}) for the definition of  $n_{\rm H, fil}$). Note that these quantities do not include the 
   gas belonging to satellite galaxies. Grey solid line corresponds to the total fraction of angular 
   momentum of gas belonging to the host halo. Cold, dense filamentary accretion accounts for 
   most of the angular momentum in the halo.}
   \label{fig:jz}
\end{figure}

Using a couple of cosmological hydrodynamics resimulations, one of a bulge-dominated galaxy 
and another of a disk-dominated galaxy at redshift 0, \citet{zavala08} show that the specific angular 
momentum of the disc galaxy closely follows that of its dark matter halo. They argue that this is 
because the stellar feedback in the simulation which produces the disk-dominated galaxy is strong 
enough to heat the gas within the halo thereby preventing it from cooling too much at early times, 
fragmenting into sub-galactic clumps, and transferring its angular momentum to the dark matter 
halo by dynamical friction. Instead this hot gas accumulates in the halo where it acquires the 
specific angular momentum of the dark matter before it eventually cools and re-collapses onto the 
central galactic disc, all the while conserving its angular momentum. However, in the first 
cosmological resimulation where individual supernovae remnants are resolved, very little gas is 
heated by stellar feedback at high redshift \citep{powell11}. Moreover, these authors show that the 
cold filaments which supply most of the gas to the galaxy 
\citep{keres05,keres09,dekel06,ocvirk08,brooks09} are not disrupted by supernova feedback. This 
result has been independently confirmed by \citet{faucher-giguere11b} who used the constant 
velocity wind model of \citet{springel03} to model feedback and found that at high redshift ($z\geq 
3$)  their supernova-driven galactic winds with low mass loading ($\eta=1$) were not able to 
suppress the accretion of cold, dense gas in intermediate-mass haloes (\mvir $\sim$ 10$^{11-12}$ 
\msun). Only when they assumed extreme parameters for their wind (i.e. loaded them with double 
the mass that is turned into stars, and drove them with double the amount of energy than that 
available from their supernova explosions) were they able to shut down cold, dense accretion. 
However this took place at the expense of the baryonic mass function which ended up dramatically 
undershooting the observations of e.g. \citet{bell03} at $z=0$. Purely based on energetic grounds, 
feedback from active galactic nuclei may be able to blast the filamentary structure 
\citep{van-de-voort10}, but it is unclear how geometrical effects will affect their capacity to do so. A 
collimated jet for instance will not, in general, deposit enough of this energy in the vicinity of the 
filaments because of their small covering factor (a few percent, \citealt{kimm11}). In any case, in 
the simulation presented here, and contrary to the conclusions of \citet{zavala08}, the specific 
angular momentum of the gas {\em in the galaxy} (thin blue line in the left panel of Fig.~\ref{fig:jtot}) 
is similar (slightly larger) to that in its DM halo host, even though supernova feedback is moderate. 
This raises the question of the importance of the numerical technique employed to assess the 
fragmentation and transfer of angular momentum in disks. Although this is beyond the scope of this 
paper, as we are chiefly interested in the outer parts of the halo, we believe that our simulation 
provides a more correct answer simply because we have better resolution. We refer the interested 
reader to the recent paper by \citet{commercon08} for a thorough comparative study between
SPH and AMR as to how a 
high artificial viscosity and numerical noise undermine the capacity of the SPH technique to 
conserve angular momentum of disks when the number of particles (total and in the kernel) is not 
high enough.

\begin{figure*}
   \centering
   \includegraphics[width=8.5cm]{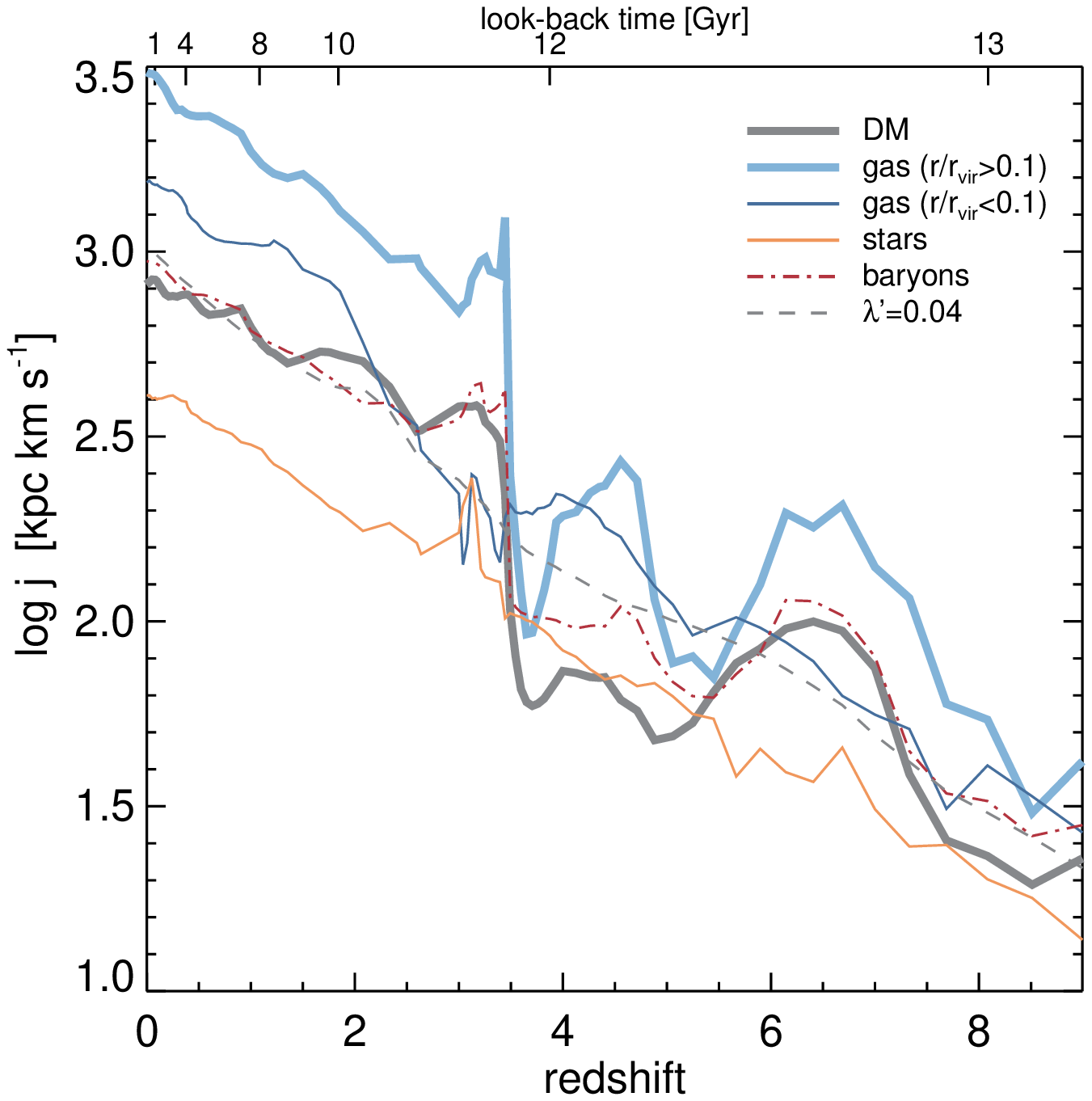} 
      \includegraphics[width=8.5cm]{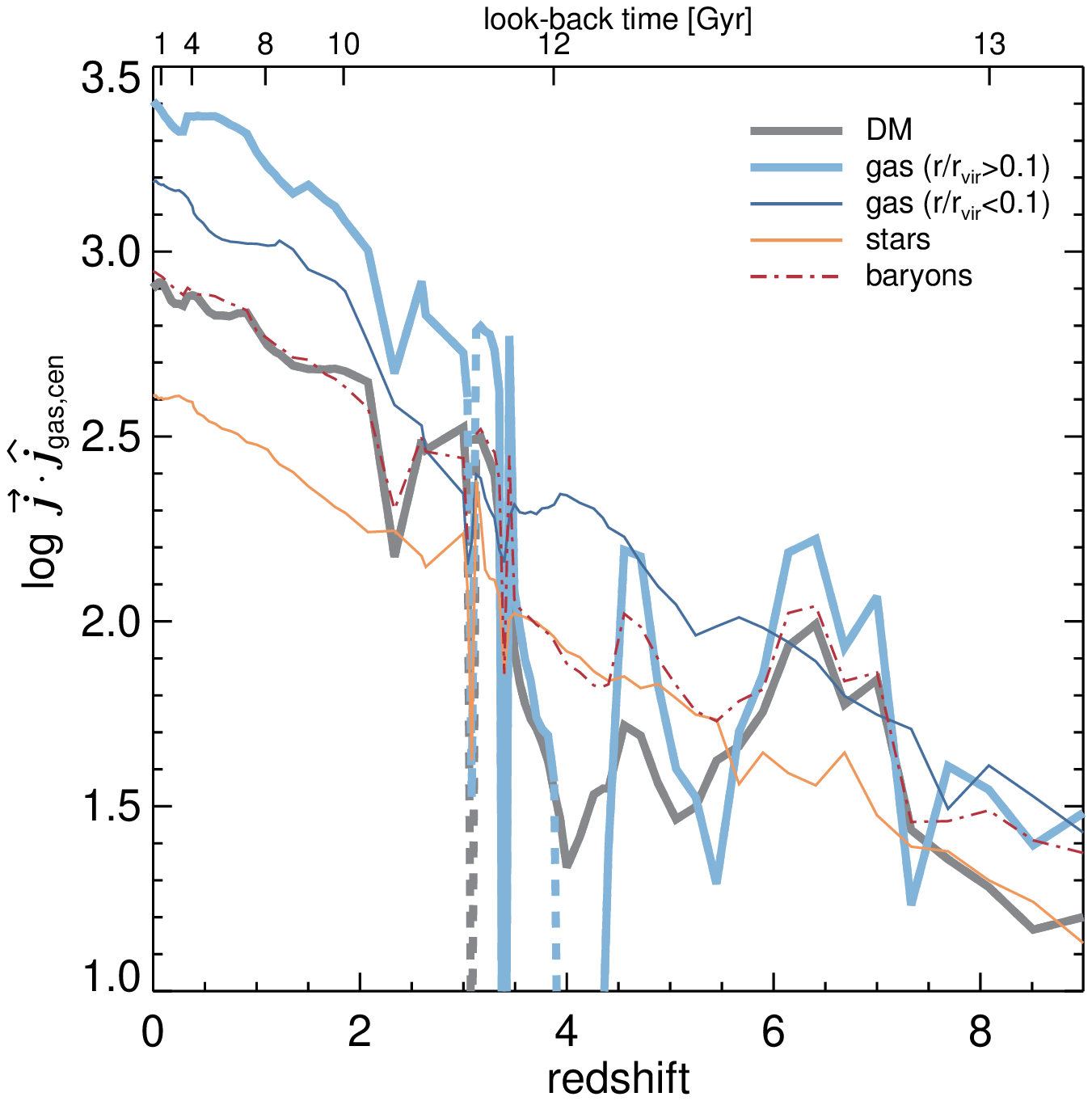} 
   \caption{Same as Fig.~\ref{fig:jtot}, but for the NutCO run. Different colours denote different 
   components: dark matter (thick grey), gas at $0.1\leq r/\rvir \leq 1$ (thick blue), central gas 
   ($r/\rvir\leq0.1$, thin blue), and stars (orange). Note that the dark matter halo and the halo 
   gas includes the contribution from satellites, while they are excluded for the measurement 
   of \jtot\ of stars. We also display \jtot\ of all baryons (star+gas) as a red dot-dashed line. 
   Specific angular momentum corresponding to $\lambda'=0.04$ is included as a grey dashed 
   line. At low redshift, the misalignment between the central gas and each component is small, 
   and gas in the outer region shows larger angular momentum than dark matter in both panels.
   }
   \label{fig:jtot_Co}
\end{figure*}

Since accretion through cold, dense filamentary gas is thought to dominate at high redshift 
\citep[e.g.][]{brooks09,powell11},  the larger \jgasout\ relative to \jdm\ is also expected to be 
associated with the dense filamentary component. To test this idea, as well as assess the 
importance of supernova feedback, we divide the gas into four different phases according to 
temperature and density as
\begin{enumerate}
\item ~~cold dense ~~($T<10^5$K and $n_{\rm H, fil}\leq n_{\rm H} < n_{\rm th}$)
\item ~cold diffuse ($T<10^5$K and $n_{\rm H} < n_{\rm H, fil}$)
\item hot  dense ~~~~($T\ge10^5$K and $n_{\rm H, fil}\leq n_{\rm H} < n_{\rm th}$)
\item hot diffuse  ~~($T\ge10^5$K and $n_{\rm H} < n_{\rm H, fil}$).
\end{enumerate}
where we define the cold dense filamentary structure using the $n_{\rm H, fil} \leq n_{\rm H} < 
n_{\rm H, th}$ gas density cut. The lower density bound ($n_{\rm H, fil}$) is chosen as
\begin{equation}
n_{\rm H,fil}\equiv \delta_{\rm f}\bar{\rho} f_{\rm bar}X_{\rm H}/m_{\rm H},
\label{eq:deffil}
\end{equation}
where $f_{\rm bar}$, $X_{\rm H}$, and $m_{\rm H}$ are the universal baryon fraction 
($\Omega_{\rm b}/\Omega_{\rm m}$), the primordial hydrogen mass fraction and the mass of the 
hydrogen atom, respectively. The dependence of the lower bound of the filament density on the background 
density ($\bar{\rho} (z)$) is motivated by the fact that the filamentary structure acquires its 
properties on large scales which are not gravitationally decoupled from the expansion of the 
Universe. The parameter $\delta_{\rm f}$  determines the overdensity of the filamentary gas. We 
find that $\delta_{\rm f}\sim100$ reasonably identifies the filamentary structure in the \nut\ 
simulations, which corresponds to hydrogen number densities higher than $n_{\rm H}\simeq0.02$ 
and $\simeq 0.001$ cm$^{-3}$ at z=9 and z=3, respectively. Note that, although this is not a worry 
in the \nut\ simulations because we spatially resolve the filaments, $\delta_{\rm f}$ will generally 
be resolution-dependent in cosmological simulations. In a rather obvious fashion, a lack of spatial 
resolution will artificially broaden the filamentary structure and thus reduce its density as mass 
needs to be conserved. 

Fig.~\ref{fig:jz} shows the contributions from the four different phases to the {\em total} angular 
momentum of the gas located at $0.1\leq r/\rvir \leq 1$. We emphasize that gas belonging to 
satellite galaxies is excluded from the measurements. It is clear from this figure that cold 
filamentary accretion is primarily responsible for the larger \jgasout. The cold filamentary gas alone 
accounts for more than half of the gas angular momentum in the region. Thus, it can safely be 
concluded that the cold, dense filamentary accretion carries material with larger \jtot\ than the dark 
matter halo from outside the virial radius to the inner region of haloes without the gas having much 
time at all to interact with the dark matter. Gas belonging to satellite galaxies occasionally 
contributes a significant fraction of  \Jtot\ (difference between the solid grey line and a horizontal 
line drawn for $\vec{\mathbf J} \cdot \hat{j}_{\rm gas,out} /{\rm J_{\rm gas,out}}=1$ in Fig \ref{fig:jz}). 
However, its impact on the specific angular momentum is less significant than that of the cold, 
dense filamentary accretion at $z\ge3$ for two reasons. First, whilst cold, dense filamentary gas 
flows into the central gaseous disc rapidly, gas gravitationally bound to a satellite galaxy orbits 
around it for the time it takes dynamical friction to drag the satellite galaxy down. As a consequence 
of this process, satellite gas angular momentum in Fig.~\ref{fig:jz} cannot be directly converted into 
the actual angular momentum of gas accreted onto the central galaxy. Second, satellites are 
accreted along the gas filaments, and hence part of the gas which we conservatively assigned to 
these satellite galaxies could be regarded as filamentary gas. In this sense, the estimate of \Jtot\ for 
the satellites in Fig.~\ref{fig:jz} should be considered as an upper bound, and thus the 
actual fraction of the total angular momentum contributed by cold filamentary accretion may be 
slightly higher than reported in this work. Finally, the contribution to the total angular momentum 
from other phases (hot diffuse, hot dense, and cold diffuse) to \Jtot\  turns out to be minor for a DM 
halo of this mass in this redshift range. Whilst the lack of angular momentum of the cold diffuse 
phase originates from its almost isotropic accretion which cancels its angular momentum to a high 
degree,  that of the hot gas phase is mainly caused by the small mass fraction of hot gas. This 
result is not very surprising because our DM halo, at $10^{11}\msun$ is not massive enough to 
sustain a stable virial shock \citep{birnboim03,keres05}, and supernova explosions in simulations 
are notoriously inefficient at ejecting large amounts of hot gas in the halo 
\citep{maclow99,dubois08,powell11,faucher-giguere11b}.

\subsection{Shock heating and the evolution of angular momentum at low redshift}

\begin{figure}
   \centering
         \includegraphics[width=8.5cm]{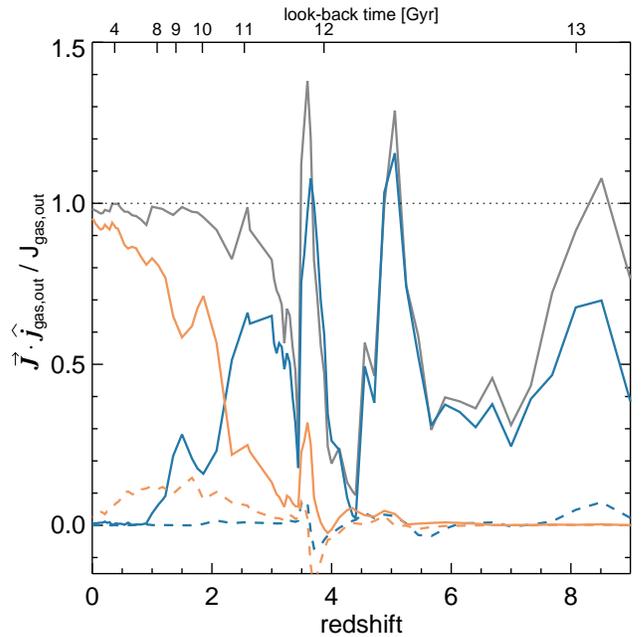} 
   \caption{Same as Fig.~\ref{fig:jz}, but for the NutCO run. The hot gas phase (orange lines) 
   begins to develop at $z\sim3$, and accounts for almost all of the angular momentum in 
   $0.1 \leq r/\rvir \leq 1$ by $z=0$. }
   \label{fig:jz_Co}
\end{figure}

Gas infall at low redshift is dominated by smooth accretion rather than mergers 
\citep[e.g.][ and references therein]{fakhouri10}. As discussed by \citet{pichon11}, this substantially 
increases the specific angular momentum of gas in the halo by delivering it coherently via the 
large-scale cosmic web. This is illustrated in Fig.~\ref{fig:jtot_Co}, where the time evolution of 
\jgasout\ for the Milky Way-like halo in the NutCO\footnote{Run with identical initial conditions and 
mass resolution to the NutFB run discussed in the previous subsection, but with lower (48 pc)
spatial resolution and no supernova feedback} simulation is found to be smoother at low redshift 
($z < 3$) than at high redshift ($z > 3$). The final mass of the dark matter halo in the NutCO run is 
$\mvir\simeq 4\times 10^{11}\msun$ (see Appendix for a time evolution of the halo mass), which is 
thought to be close to the mass at which gas accretion transitions from the cold mode to the hot 
mode \citep{birnboim03,ocvirk08}. Since the NutCO run does not include supernova feedback, 
satellite galaxies can retain more gas than in the NutFB run, and thus the resulting relative 
contribution from the cold, dense filamentary gas to the total angular momentum gets slightly 
smaller at high redshift than the feedback run (Fig.~\ref{fig:jz_Co}). However, there is little difference 
in \jgasout\ or \jdm\ between these two runs for  $z \geq 3$. Fig.~\ref{fig:jz_Co} also shows that at 
$z\lesssim2$, the diffuse hot gas phase becomes the dominant reservoir of angular momentum in 
the region ($0.1\leq r/\rvir \leq1$). Considering that the post-shock temperature for an isothermal 
sphere of mass a few times $10^{11}\msun$ (our halo reaches 2  $\times 10^{11}\msun$ at $z 
\simeq 2$, c.f. Appendix A) is $ T_{\rm shocked} \gtrsim 3 T_{\rm vir}/8 \sim 10^5 {\rm K}$ 
\citep{dekel06}, where $T_{\rm vir}(\simeq35.9 \times (V_{\rm c}/[\rm{km \cdot s}^{-1}])^2)$ is the 
virial temperature, and that the run does not feature any SN feedback, we can safely identify the 
halo hot gas at $z\lesssim 2$ with material which has been shock-heated. Nonetheless, as 
Fig.~\ref{fig:jtot_Co} clearly demonstrates,  this hot halo gas within $0.1 \leq r/\rvir \leq 1$ still has 
a specific angular momentum higher than that of the dark matter, indicating that shock-heating of 
the accreted cold gas, even when it takes place as early as $z=2$, does not erase the difference in 
specific angular momentum of dark matter and gas.

\begin{figure}
   \centering
         \includegraphics[width=8.5cm]{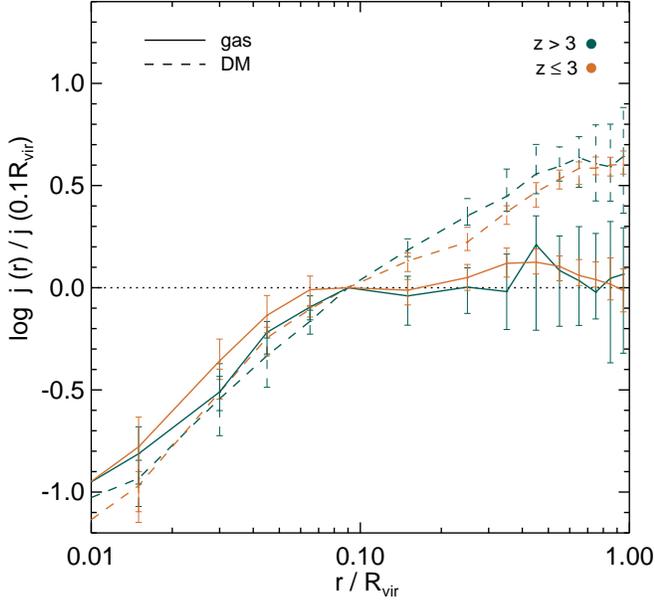} 
   \caption{Stacked distributions of normalised specific angular momentum as a function of radius 
   from the NutCO run. The specific angular momentum of gas and dark matter is shown as solid 
   and dashed lines. Blue lines display \jtot\ at high redshift ($z>3$), while orange lines show \jtot\ 
   at low redshift ($z\leq3$). Each point  corresponds to the median, and interquartile ranges are 
   included as error bars. The normalised \jtot\ of gas is flat at $0.1 \lesssim r/\rvir \lesssim 1$, 
   and gets diminished in the region where the galactic disc is located, suggesting that \jtot\ of gas 
   is reasonably conserved  up to $r=0.1\rvir$.}
   \label{fig:j_rstat}
\end{figure}

As is the case in the NutFB run, we find that \jbar\ is comparable to \jdm\ up to z=0, whilst 
\jgasout\ is always greater than \jdm. Once again, given that gas accretion at the virial radius of the 
halo is never perfectly coplanar with the galactic disc, we expect the (small) off-axis components 
($j_x$ and $j_y$) to vanish before gas becomes part of the central galaxy, and thus the modulus of 
gas specific angular momentum not to be conserved in general. We also expect gas accretion history 
to reduce the modulus and smooth the time evolution of \jgascen with respect to \jgasout. Fig.~\ref{fig:j_rstat} 
shows the radial profiles of  the \jtot\ distributions of gas and dark matter within the 
virial radius of the dark matter halo, normalised to their respective values of  \jtot\ at 0.1\rvir. These 
profiles are split in a low and  a high redshift bin ($ z\leq 3$ and $z > 3$) and stacked. Two striking 
features emerge from the analysis of Fig.~\ref{fig:j_rstat}. First the dependence of the shape of {\em 
all} the specific angular momentum profiles on redshift is minor. Second, while the \jtot\ of the dark 
matter component drops almost proportionally with radius, the \jtot\  profile of the gas is flat in the 
outer parts of the halo ($r \geq 0.1$ \rvir). Although we did not included the curve in this paper, 
this is in stark contrast with the behaviour of the gas specific momentum in the NutAD run 
which closely follow the $j$ profile of the dark matter halo.
This implies that as long as the gas can radiatively cool, its 
specific angular momentum is conserved whilst  it is being advected to the inner region ($r < 0.1$ 
\rvir), regardless of whether shock heating occurs or not. Therefore, any redistribution of angular 
momentum must happen within the central region of the halo, close to the galactic disc. The detailed 
analysis of this complex process is beyond the scope of the current paper and will be presented 
elsewhere  (Tillson et al. {\sl in prep.}).

\subsection{Generalisation of the results}
\label{sec:gen}

Now that we have established that gas in a Milky Way-like halo has a larger specific angular 
momentum than dark matter, it is vital to know whether this property is generic  or not. To build a 
statistically representative sample of haloes, we make use of three large cosmological 
hydrodynamics simulations called \mn, Cosmo25, and Cosmo50. The \mn\ simulation contains 326, 
1921, 3307, and 4015 haloes with $\mvir \ge 10^{11}\msun$ at z=6.1, 3.8, 2.5, and 1.5. Since the 
simulation has only  been carried out down to z=1.5, we use the Cosmo25 and Cosmo50 run to 
supplement the sample at lower redshifts. The number of haloes with $\mvir \ge 10^{11}\msun$ in 
the Cosm25 run is 378 and 459 at z=1.5 and z=0 respectively. For the Cosmo50 run, we restrict 
our analysis to 481 more massive haloes ($\mvir \ge 10^{12}\msun$) at z=0 because of the poorer 
resolution of the simulation.

\begin{figure}
   \centering
   \includegraphics[width=8.5cm]{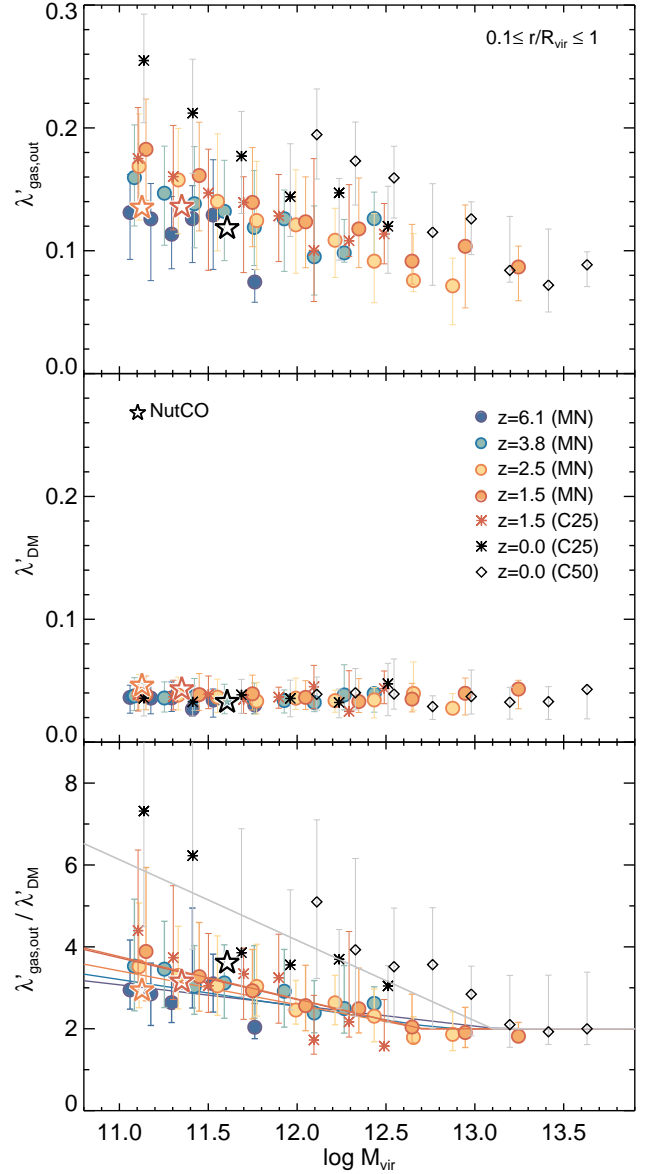} 
   \caption{Spin parameter distributions of the gas within $0.1\leq r/\rvir \leq 1$ (top), the dark matter 
   halo (middle), and the ratio of the two (bottom). Filled circles correspond to the results from the 
   \mn\ simulation, and asterisks and diamonds show the results from the Cosmo25 and Cosmo50 
   run, respectively. Different colours denote different redshifts as indicated on the figure. The 
   results from the NutCO run are also included as stars with the same colour-coding. Solid lines 
   show linear fits to the simulated data at different redshifts as given by Equation~(\ref{eq:fitratio}). 
   Statistically, the specific angular momentum of the gas within $0.1\leq r/\rvir \leq 1$ is at least 
   twice that of the dark matter halo host, at all redshifts.}
   \label{fig:lamratio}
\end{figure}

The middle panel of Fig.~\ref{fig:lamratio} shows the median and the interquartile range of the spin 
parameter distribution of the dark matter haloes. As is well known in the literature, it shows that the 
typical value for the spin is $\lambda\simeq 0.04$ with very little dependence on halo mass and 
redshift. On the other hand, the spin parameter of gas in the \mn\ simulation (filled circles) is 2--4 
times higher in general than that of the dark matter halo and clearly depends on halo mass. 
The ratio turns out to be even more significant at z=0 in the Cosmo25 (asterisks) and Cosmo50 run 
(diamond), demonstrating that our finding that gas has larger specific angular momentum extends 
to {\em all} halos over the {\em entire} redshift range. Moreover, it also substantiates our claim that 
gas brought in later in the life of the halo has more specific angular momentum because of  the 
large scale origin of this angular momentum which has been investigated in a companion paper 
\citep{pichon11}. Note that since the \mn\ and Cosmo25 simulations are run using different 
cosmologies (WMAP1 and WMAP5 respectively) and feature different mass resolutions, we 
included data at  a common redshift ($z=1.5$) to ensure that the differences between the z=1.5 
\mn\ and the z=0 Cosmo25 halos does not arise from a different choice of cosmology/simulation 
parameters.

For practical purposes, we compute a redshift-dependent fit to the specific angular momentum gas 
to dark matter ratio for halos with $\mvir \leq 10^{13}\msun$ 
{\small
\begin{equation}
\frac{\lambda'_{\rm gas,out}}{\lambda'_{\rm DM}}\simeq \max[
         3.5+24.3(1+z)^{-0.8}~-1.97(1+z)^{-0.7} \log \mvir ,2].
         \label{eq:fitratio}
\end{equation}
}
We emphasize that all the extra specific angular momentum brought in by the gas will not be 
devoted to spinning up the central galactic disc, given that its typical misalignment with the central 
gaseous disc is measured to be $\approx 40^{\circ}$  for intermediate mass halos 
($\mvir\sim10^{11}\msun$) in the \mn\ simulation. Moreover, in the vicinity of the central disc, 
hydrodynamical interactions with the circum-galactic gas will become important and may 
substantially redistribute angular momentum. A careful numerical investigation of angular 
momentum advection in the central region is therefore needed to determine whether or not a 
simple semi-analytic approach is capable of correctly describing the evolution of angular 
momentum of disc galaxies using the new initial conditions we provide with this fit.

We also note that limited spatial resolution of our large volume cosmological simulations will lead to 
an artificial increase of the gas angular momentum, especially in low mass halos.
Indeed, the softening of the gravitational force produces very extended discs at the centre of these halos, 
which would probably be contained within 0.1 \rvir\ at higher resolution. However, 
comparing the spin of the gas in the NutCO halo (large empty stars in Fig. \ref{fig:lamratio}) at various 
redshifts with that measured for a sample of halos of comparable mass available in these 
cosmological simulations (solid circles and asterisks in Fig. \ref{fig:lamratio}), 
we conclude that resolution effects most likely account for a minor fraction of the gas spin ($\sim20\%$).

\subsection{Comparison with observations}

\begin{figure}
   \centering
   \includegraphics[width=8.5cm]{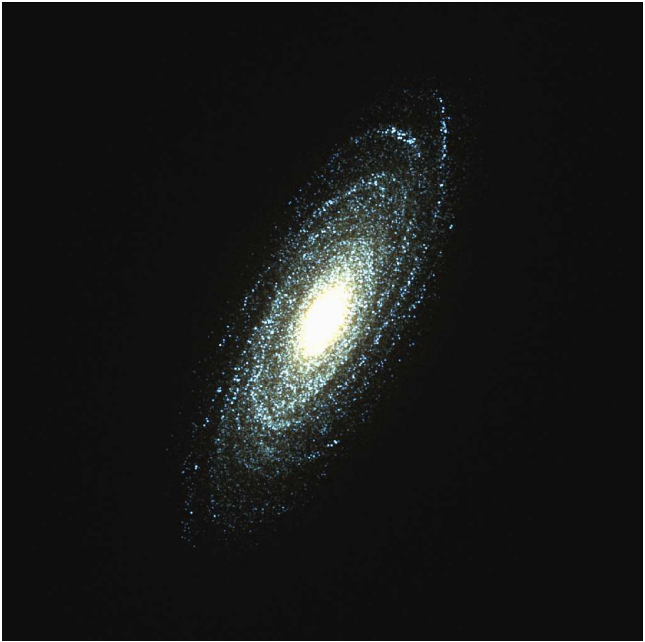} 
   \caption{True colour composite image of the NutCO galaxy in the U, B and V at $z=0$.
   The image measures 15 kpc on a side. Note that internal extinction by dust is not accounted for in this image. 
   Notice the presence of a large bulge (I-band B/D ratio 0.78) typical of a galaxy with morphological type 
   Sa/b (see text for detail).}
   \label{fig:galimage}
\end{figure}

Ultimately, to decide whether or not our simulations  yield realistic galaxies, we have to compare 
their properties with observations. Unfortunately, with the high resolution we need to properly 
address the issues related to angular momentum, cooling and feedback, we can only perform 
such a comparison at $z=0$ with one of our runs (NutCO) at the time being. As this simulation 
does not model any feedback mechanism, by z=0 the simulated galaxy has formed too many stars ($M_{\rm star}=6.8\times10^{10}\msun$). 
At $z=0$, the $I$-band bulge-to-disc ratio ($B/D$) of the central galaxy 
in the NutCO run is $\simeq 0.78$ (see Fig.~\ref{fig:galimage}). Such a value is close to the 
typical $B/D$ of Sa/Sb type spirals, but the disc scale length (2.0 kpc) turns out to be smaller 
than those observed \citep{graham08}.

 \begin{figure}
   \centering
   \includegraphics[width=8.5cm]{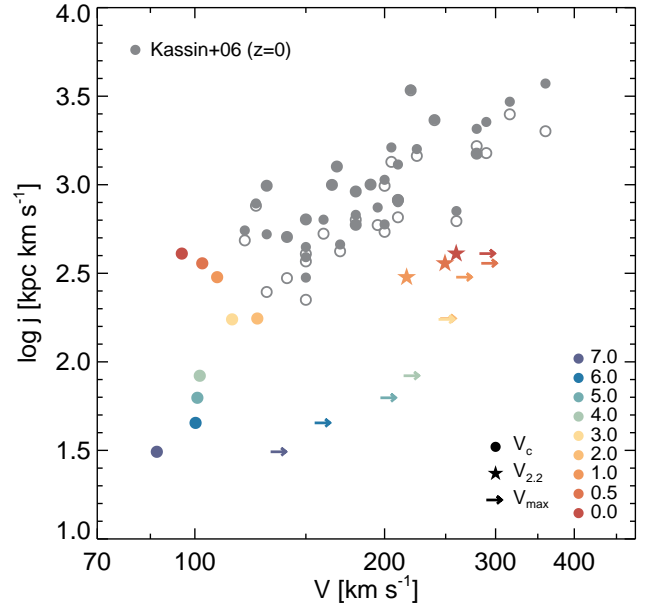} 
   \caption{Comparison between the NutCO galaxy and the observational bright disc galaxy sample 
   described in \citet{kassin06}.
   Observational data is in grey (lower and upper estimates of $j$ are given by empty and solid 
   circles respectively, see Kassin et al. 2011 for details). For the simulation data, different symbols 
   (circles and right arrows) stand for different estimates of the velocity of the galaxy 
   (circular velocity and maximum rotation velocity). We also include estimates of $V_{\rm 2.2}$,
    the velocity measured at 2.2 times the exponential scale length in $I$ band, for $z\leq 1$. 
   Note that the simulated galaxy displays a dense large bulge which overgrows, at least partially, 
   because of the lack of modelling of any kind of internal feedback (radiation, stellar winds, 
   supernovae, cosmic rays, etc.) in the NutCO run.}
   \label{fig:comp_obs}
\end{figure}

Nevertheless, bearing this caveat in mind, we present in Fig.~\ref{fig:comp_obs} measurements of 
the $j$ versus $V$ relation for the NutCO galaxy at various redshifts, compared to the 
observational bright disc galaxy sample gathered by \citet{kassin06} (at $z \simeq 0$). 
Kassin et al. derived the specific angular momenta for these galaxies from Halpha + HI rotation curves 
and radial stellar mass distributions given in \citep{kassin06}.
The average errors in $j$ measured from the data are $\approx 60 \,{\rm kpc\, km s^{-1}}$. 
As is done in the observations, 
both the amount of specific angular momentum and the velocity of the simulated galaxy ($V_{\rm 
max}$, maximum of the rotation curve) are estimated using the stellar component, except for the 
filled circle symbols where the velocity is measured as the circular velocity at the virial radius of the 
dark matter halo. For a fair comparison with observations, we also include velocities measured at 
2.2 times the disc scale length in the $I$ band ($V_{\rm 2.2}$) at $z\leq1$.

 \begin{figure*}
   \centering
   \includegraphics[width=8.5cm]{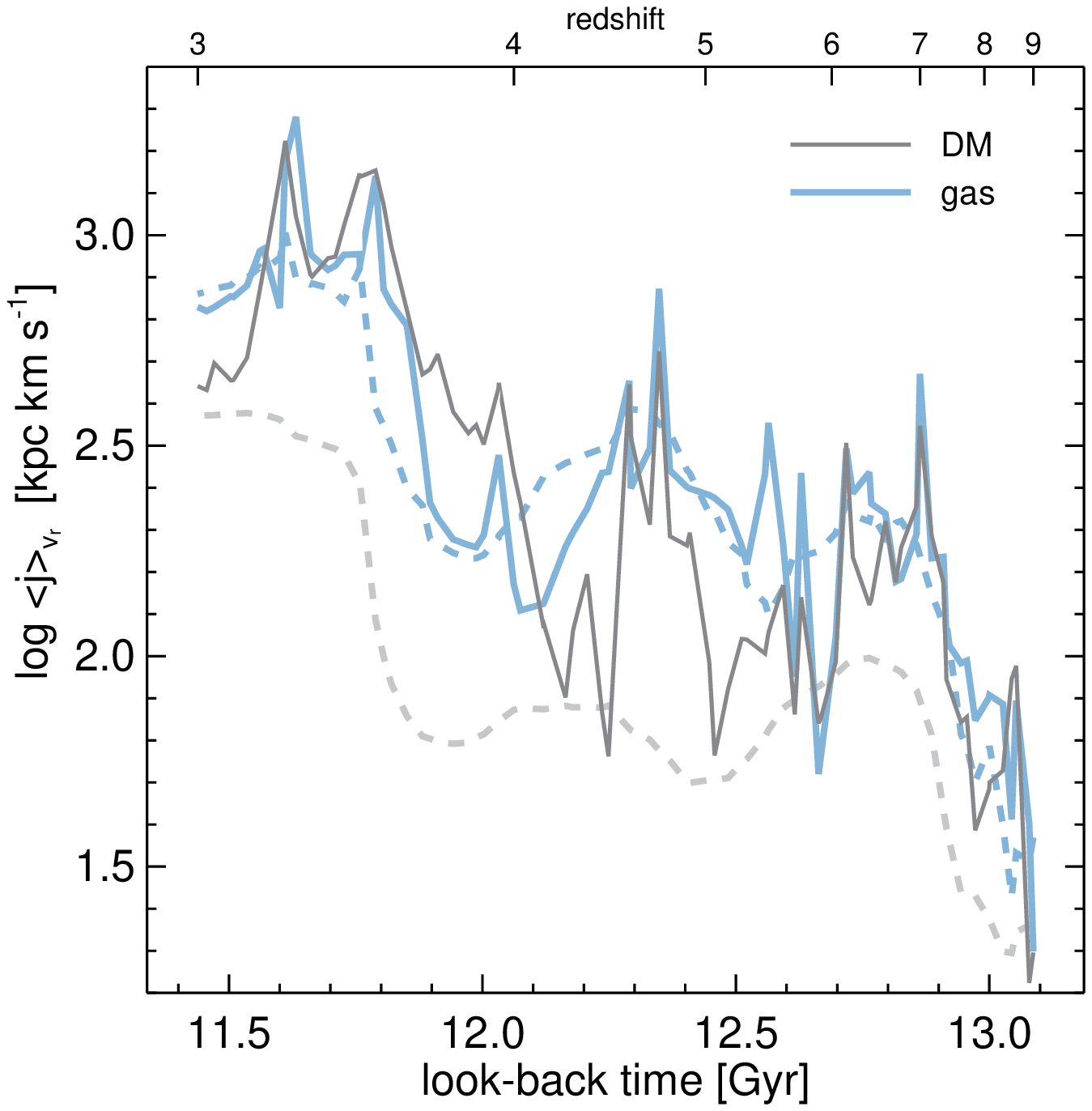} 
      \includegraphics[width=8.5cm]{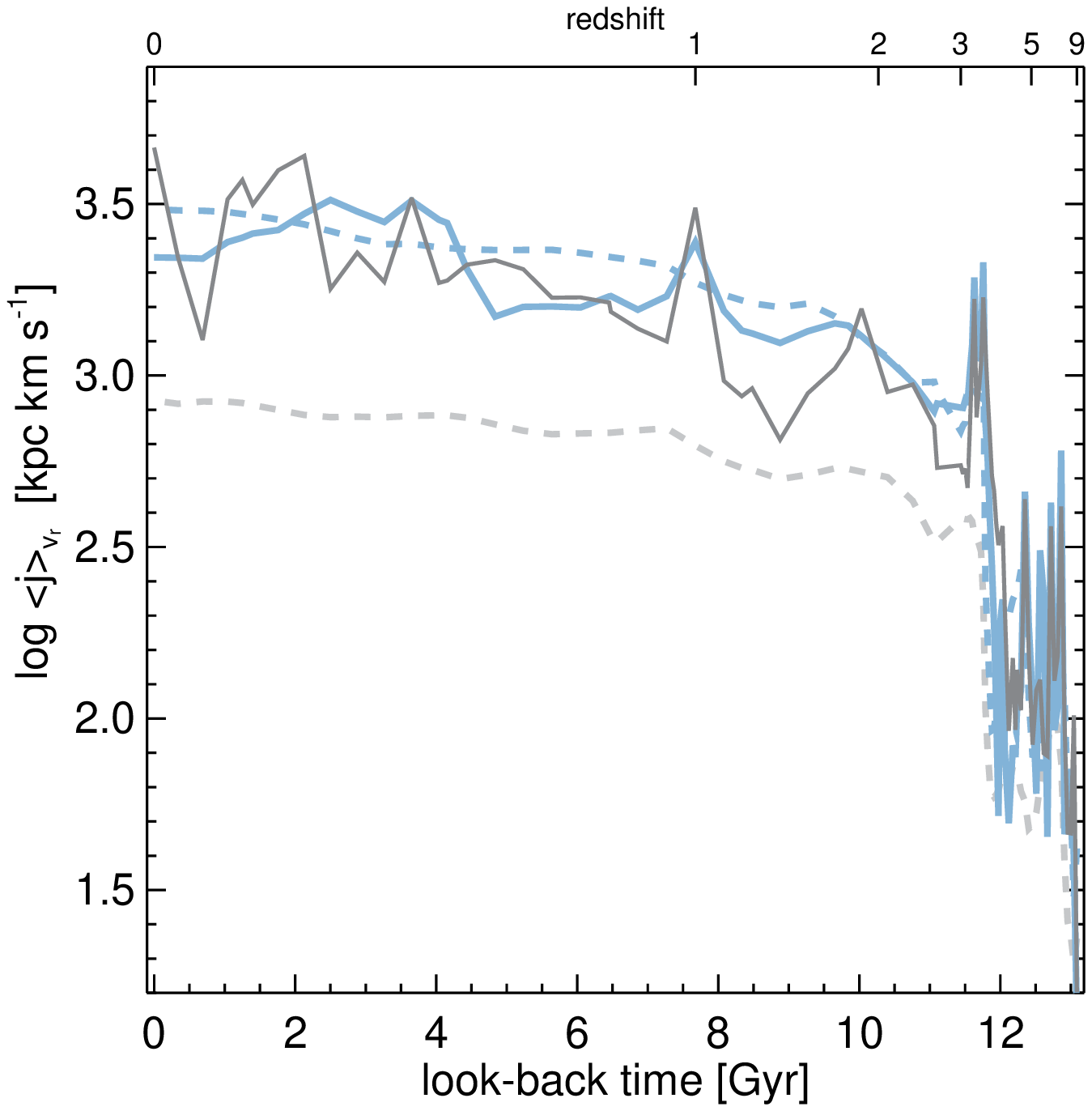} 
   \caption{Accretion-weighted specific angular momentum for dark matter (black solid line) and 
   gas (blue solid line) at the virial radius in the NutFB (left panel) and the NutCO (right panel) runs. 
   The specific angular momenta of the dark matter and the halo gas (\jgasout) 
   shown in Fig.~\ref{fig:jtot} and \ref{fig:jtot_Co} are included as grey dashed and blue dashed lines. 
   Note the similar evolution of the specific angular momentum of the halo gas and of both accreted gas 
   and dark matter. }
   \label{fig:jratio}
\end{figure*}

Looking at Fig.~\ref{fig:comp_obs}, one clearly sees that the velocity estimate plays a crucial role in our 
ability to assess whether simulated galaxy stars have the correct amount of angular momentum. This is 
because the observed slope of the relation ($\approx$ 2) is quite large so any error on the velocity will 
translate into a much larger error on the specific angular momentum. More specifically, our simulation 
suffers from forming too many stars and as a result any velocity estimated in the central region of the halo 
is bound to be too large. If, on the other hand, we use the circular velocity at the virial radius of the halo to 
bypass the problem (note that this is current practice, see e.g. \citealt{dutton09}), we find that the level of 
angular momentum of the simulated stars is in fair agreement with the observations (Fig.~\ref{fig:comp_obs}). 
We point out that this is a consequence of the lossless transport of a large amount  of specific angular 
momentum by gas from super halo scales  right into the inner regions, followed by its redistribution in the 
vicinity of the disc, as discussed in the previous section. As a result of this effect and the low efficiency of 
the star formation process on disc scales, stars end up with a factor 2 to 3 less angular momentum than 
their dark matter halo counterpart which is very close to the discrepancy between dark matter only 
simulations and observations, as pointed out by \citet{navarro00} and Kassin et al. (2011, {\sl in prep.}).

Finally, it is interesting to note that when we use the $V_{\rm max}$ estimates for the velocity of the 
galaxy, although we do not match the zero point of the relation, simulated data points at different redshifts 
move along the observed sequence, suggesting that we would predict very little time evolution of the 
sequence. On the contrary, $V_{\rm c}$ estimates suggest there exists a strong evolution of the relation 
with redshift as the flat velocity profile of the dark matter halo is already in place very early on.

\section{Origin of the discrepancy between dark matter and gas specific angular momenta}
\label{sec:diff}

Having robustly demonstrated that the amount of specific angular momentum carried by gas and dark 
matter differ throughout the virialised halo and at all redshifts, we now turn to identifying the cause of 
such a difference in \jtot.  There are essentially three possibilities: i) the gas carries a larger amount of 
specific angular momentum than the dark matter at the time of accretion, or ii) the dark matter component 
does not conserve \jtot\ as well as the gas or iii) the difference in \jtot\ reflects a different spatial 
re-distribution of the angular momentum of gas and dark matter within the virialised halo.

In order to test the first possibility, i.e. whether the gas already carries a larger amount of specific angular 
momentum than the dark matter when it is accreted by the halo, we compute the accretion-weighted 
specific angular momentum modulus of both components at the virial radius as 
\begin{equation}
\left<j\right>_{v_r} = \left| \frac{\sum_i m_i v_r \Theta(-v_r) \vec{\mathbf{r}}_i \times \vec{\mathbf{v}}_i}{\sum_i m_i v_r \Theta(-v_r)} \right|,
\end{equation} 
where $\Theta$ is the Heaviside step function. This is done by calculating the contribution of infalling 
($v_r<0$)  gas  or DM particles located within the thin shell defined by $0.95 \leq r/\rvir \leq 1.05$. Since 
the collisionless dark matter particles can potentially get accreted several times as they come in and out of 
the halo, we track each particle individually and flag them the first time they cross the virial sphere so as 
not to re-accrete them later. For the collisional gas, this problem does not arise as it remains in the center 
of the halo unless blown out/evaporated by feedback or shock-heating, so a simple radial velocity cut 
suffices.

Fig.~\ref{fig:jratio} shows that the specific angular momenta of the accreted gas and dark matter agree 
reasonably well  both in the NutFB and NutCO run, at all times. Although we have not included it in this 
paper, this turns out to be the same in the NutAD run. It is noteworthy that both freshly accreted gas and 
dark matter bring in a larger amount of  specific angular momentum than that of  the dark matter halo, 
\jdm, which is  the common assumption used in (semi-)analytic studies. The fact that \jgasout\ carries 
nearly the same large amount of specific angular momentum as $\left<j\right>_{v_r}$ supports the view 
that the angular momentum of the gas accreted by the central galaxy is acquired {\em before} it enters the 
dark matter halo. As argued elsewhere \citep{pichon11}, this favours the large-scale motion of the cosmic 
web as the most likely origin of the angular momentum of galaxies.

The second possibility, i.e. that the amplitude of the dark matter component specific angular momentum 
is conserved to a lesser degree than the gas, is easily ruled out on the grounds that, 
as argued in Section \ref{sec:highz}  when the time evolution of the total specific angular momentum 
of {\em all} the baryons is calculated, it matches that of \jdm\ quite accurately. It seems therefore 
impossible to argue that dark matter loses more specific angular momentum than the gas as the main 
reason why \jgasout\  is larger than \jdm.

\begin{figure}
   \centering
   \includegraphics[width=8.5cm]{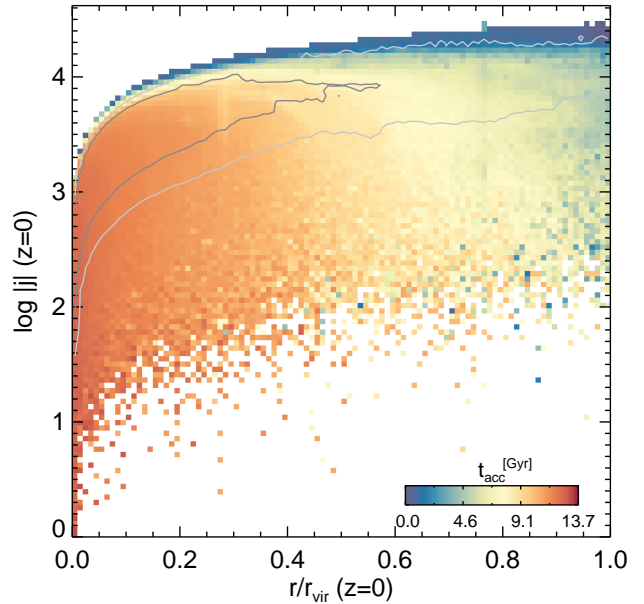} 
   \caption{Distribution of angular momentum of DM particles as a function of their position for the 
   virialised DM halo in the NutCO run at $z=0$. Colors indicate the lookback time at which particles 
   were accreted for the first time. Dark and light grey contours enclose 50 and 90 percent of the mass 
   respectively. Note the large range of angular momentum values spanned by individual particles at any 
   radius. }
   \label{fig:dmpos}
\end{figure}

We are then left with the third possibility: the dissipative nature of the gas leads to a transfer of specific 
angular momentum distinct from that of the dark matter. As previously mentioned, cold, dense filamentary 
gas flows into the inner part of the dark matter halo in a dynamical time without undergoing shock-heating 
\citep[e.g.][]{brooks09,powell11}. The gas subsequently settles onto the central gaseous disc, and, as a 
result,  the outer region ($0.1\leq r/\rvir \leq 1$) is  occupied by gas recently accreted, which has a large 
amount of specific angular momentum. On the other hand, dark matter particles pass through the central 
region of the halo and depending on their exact orbital properties, can end up populating the outer regions 
despite having entered the halo at earlier times with smaller angular momentum. This is shown in 
Fig.~\ref{fig:dmpos} where we plot the angular momentum as a function of position for all the DM particles 
belonging to the NutCO halo at z=0, colour encoding the time at which they where accreted. Whilst a lot 
of particles in the outer regions were accreted  fairly recently (less than 5 Gyr ago), there exists a non-
negligible fraction of them which were accreted at much earlier epochs (more than 7 Gyrs ago).  Thus, 
despite the fact that dark matter first falls into the halo with a similar amount of specific angular 
momentum as the gas (Fig.~\ref{fig:jratio}),  mixing with `older' particles makes \jdm\ become 
systematically smaller than \jgasout. Note that this also explains the similar evolution of the specific angular momentum of the halo gas (\jgasout)  and the newly accreted gas at high redshift (Fig.~\ref{fig:jratio}).

 Obviously the cold, dense filamentary gas accretion disappears at low redshift  
\citep[e.g.][]{kimm11,faucher-giguere11a,stewart10},  and therefore one expects that the high redshift 
segregation of specific angular momentum between gas and dark matter previously advocated becomes 
weaker as time elapses. Indeed, in the extreme case where gas is prevented from cooling (NutAD run: 
Fig.~\ref{fig:jtot_ad}) the evolution of its specific angular momentum closely  follows that of the dark 
matter, indicative of a similar degree of mixing of low and high specific angular momentum material for 
both components. However, when gas can radiatively cool, one expects this mixing process to be much 
less efficient as it will generally take place on timescales (typically a few sound crossing times of the 
halo) which are longer than the cooling time of the hot central gas. This means that the shock-heated 
material at the centre of the halo, which has a lower specific angular momentum than that in the outer 
regions, is cooled and accreted onto the central disk before it can mix with higher specific angular 
momentum gas. This creates a `cooling flow' towards the central galaxy and as a result preserves the 
distribution of gas specific angular momentum that was set in at high redshift (Figs.~\ref{fig:jtot_Co} 
and \ref{fig:j_rstat}). Note that this also explains the increasing discrepancy between dark matter and 
gas spin as the mass of the halo drops that we measured in large scale cosmological simulations 
(Section ~\ref{sec:gen}, Fig.~\ref{fig:lamratio}): as the halos become larger, cooling timescales 
increase (because of higher gas temperatures and smaller central densities) allowing more and more 
mixing between low and high specific angular momentum gas to occur. As a result, the gap between 
\jgasout\ and \jdm\ is reduced for the most massive objects. 
 
The redshift dependence of the spin parameter of the gas can also be understood in this framework. 
Again, since the cooling permits the gas to flow into the centre, \jgasout\ tends to reflect \jtot\ of newly 
accreted gas. This means that, for any dark matter halo of a given mass, its gas content at lower redshift 
will be composed of material carrying larger specific angular momentum. In other words, the redshift 
dependence of gas spin shown in Fig.~\ref{fig:lamratio} is understood as a large scale structure driven 
increase in the specific angular momentum of freshly accreted gas.  

Even though we do not investigate what causes this behaviour in this paper, we remark that the systematic 
discrepancy between \jdm\ and $\left<j\right>_{v_r}$ measured in the NutCO run at {\em all} times 
means that the dark matter component of the halo (and its baryons as a whole) loses somewhere between 
half and two thirds of its specific angular momentum amplitude, depending on redshift (Fig.~\ref{fig:jratio}). 
We note that even though this result seems in agreement with the measures of \citet{book11} for a handful 
of well resolved dark 
matter halos in pure N-body simulations, we do not find, like these authors, that it can be attributed to 
external torques, following the original argument of \citet{peebles69}. The reason for this disagreement 
lies in the {\em continuous} nature of the specific angular momentum loss of the dark matter component.  
As Fig.~\ref{fig:jratio} demonstrates, the accretion weighted specific angular momentum of dark matter 
is higher than the specific angular momentum of the dark matter halo at {\em all redshifts}, regardless of 
whether we are in a phase where tidal torque theory predicts the angular momentum of the halo should 
grow or not. We are therefore pushed to conclude that the angular momentum `loss' suffered by the halo 
has to be attributed primarily to  a `vector cancellation' effect: particles with similar amounts of angular 
momentum (modulus) but pointing away in opposite directions contribute only a small amount to the `net' 
angular momentum of the virialised halo \citep[e.g.][]{vitvitska02}.  
Note that the accretion history effect will also play a role in 
reducing the modulus of ${\vec \mathbf j}_{\rm dm}$ but cannot account for all the decrease.    

Finally, the admittedly low spatial resolution (1--2 kpc) of the large scale cosmological hydrodynamics 
simulations we use to generalise our results, prevents dark matter (and gas) from forming more 
concentrated structures.  As a result, galaxies in small halos ($N_{\rm DM}\sim 2\times10^3$) show a 
more extended disc than their \nut\ series counterparts.  However, we believe that the properties of halos 
we discussed in this paper, since they are integrated over a large fraction of the halo radius, are robust. A 
caveat to bear in mind however, is that the physical processes that can prevent the cooling catastrophe in 
cluster environments are likely to change the ratio $\lambda_{\rm gas,out}/\lambda_{\rm dm}$. For 
example, as extensively discussed in the literature, feedback from active galactic nuclei is thought to stir/
heat a large amount of cooling gas, redistributing it in the outer parts of the halo. As previously argued, 
such a mixing mechanism will contribute to narrow the gap between gas and dark matter spin. Such a 
potent feedback mechanism is not included in the cosmological simulations presented here, and thus a 
larger fraction of baryons than observed are converted into stars in massive haloes \citep{dubois10}. This 
should be taken into account before concluding that the ratio 
($\lambda_{\rm gas,out}/\lambda_{\rm dm}$) for massive haloes has converged to $\sim$ 2 as 
Fig.~\ref{fig:lamratio} (bottom panel) advocates.

\section{Conclusions}

In this paper, using high-resolution cosmological hydrodynamics re-simulations of individual galaxies 
and large scale cosmological hydrodynamics simulations, we have revisited the standard theory for 
angular momentum evolution of gas and dark matter within virialised structures.  

According to this standard picture, gas acquires angular momentum through nearby tidal fields in the 
same way as dark matter and is shock-heated to the virial temperature during the collapse of the halo. 
This gas then radiatively cools and settles into a rotationally supported disc, conserving its specific 
angular momentum in the process \citep{fall80, dalcanton97, mo98}. Such assumptions naturally lead to 
predict that dark matter and halo gas share the same specific angular momentum. Investigating the issue,  
our main conclusions can be summarised as follows:
\begin{itemize} 
\item Indeed, the time evolution of the halo gas specific angular momentum is quite comparable to that of 
the dark matter, but only when gas is not permitted to radiatively cool, i.e. it is artificially forced to shock-
heat during the collapse (NutAD run).
\item When radiative cooling is turned on, the angular momentum transport of gas becomes distinctively 
different from that of dark matter (NutCO run). Even though gas and dark matter still bring in a similar 
specific angular momentum ($\left<j\right>_{v_r}$) when they are accreted, this amount turns out to be 
systematically larger than the specific angular momentum of the virialised dark matter halo as a whole 
(\jdm) by a factor 2 to 6, depending on redshift and halo mass (see also \citealt{stewart11}).
\item Regardless of whether cooling is turned on or not, the modulus of the specific angular momentum 
inside the virial sphere is not conserved for dark matter or gas. The amount lost ranges between 
one half and two thirds depending on redshift. We attribute this continuous `loss' of angular momentum 
to a `vector cancellation' effect (the angular momentum of newly accreted material is never perfectly 
aligned with the angular momentum of the whole halo) augmented by an accretion history effect 
(material accreted earlier carries less angular momentum so it weighs the average modulus of the specific 
angular momentum down at any epoch). 
\item When radiative cooling is turned on, provided one compares the dark matter with the total baryon 
(gas and stars) specific angular momentum, their amplitudes are similar at all times. In other words, we 
find no evidence of baryons losing more (or less) angular momentum than dark matter within the 
virialised halo at any epoch.
\end{itemize}

More specifically, our analysis reveals that at high redshift, the discrepancy between gas and dark 
matter specific angular momentum within $0.1 \leq r/\rvir \leq 1.0$ (\jgasout\ and \jdm, respectively) 
arises because the angular momentum-rich freshly accreted gas flows into the central region through cold, 
dense filamentary accretion without being redistributed throughout the halo by shock-heating or  
supernovae feedback (NutFB run). As the dark matter halo grows at lower redshifts, a progressively larger 
amount of gas undergoes shock-heating, but since the central dense gas can still cool and collapse onto 
the disc on timescales shorter than a halo sound crossing time, the difference between \jgasout\ and \jdm\  
is, by and large, preserved. This competition between central cooling and redistribution of gas through 
shock heating and/or feedback induces a mass dependence of the discrepancy between gas and dark 
matter specific angular momentum: it is reduced for more massive halos where shock heating dominates. 
However, for a halo of fixed virial mass, this discrepancy is also larger at low redshifts than it is at high 
redshifts because the amount of specific momentum carried by the newly accreted material grows faster 
than the halo average with time (cf. \citet{pichon11} for the origin of this effect). This generic 
behaviour is encapsulated in Equation~(\ref{eq:fitratio}), which provides a fit to our simulated data.

Our efforts to probe the spatial distribution of angular momentum within the virialised halo also led us to 
define an inner region ($ r/\rvir\leq0.1$). Gas in this region, \jgascen, generally has more specific angular 
momentum (by a factor 2 or so) than the dark matter halo as a whole, and only stars have less (also by 
about a factor 2).  We attribute this dichotomy within the baryon component to the long global star 
formation timescale of the disc compared to the gas accretion time scale, driving stars to preferentially 
form from gas accreted at a much earlier stage with lower angular momentum (see also \citet{dutton09} 
who reach similar conclusions with a SAM).  As a result, a comparison with the data at z=0 reveals that 
the level of specific angular momentum of our high resolution simulated stellar disc matches quite well 
that of the observations. However the difficulty of obtaining realistic galaxy rotation curves still prevents 
us from satisfactorily reproducing the observed $j$ versus $V$ relation (see Fig.~\ref{fig:comp_obs}). 
Part of this failure is to blame on the 
fact that the only re-simulation we have been able to run down to z=0 so far (NutCO run) does not include 
any modelling of feedback (stellar winds, supernovae, cosmic ray, etc.) \citep[as pointed out by e.g.][]{governato07,agertz11,brook11}. For this reason, too many 
baryons (the universal fraction $\Omega_B / \Omega_m$) remain in the host halo and concentrate in the 
galaxy, potentially affecting the morphology of our galaxy which features a large bulge (morphological 
type Sa/Sb) but in any case leading us to overestimate disc rotational velocities.

Overall, our results demonstrate the need for the standard picture of galaxy disc formation and evolution 
to undergo a major overhaul as the fundamental hypothesis upon which it is based (specific angular 
momentum of the gas and dark matter within the virialised halo are the same, gas specific angular 
momentum is conserved) seem to be incorrect. This should have important consequences for galaxy disc 
sizes and therefore star formation timescales estimates used in virtually all galaxy formation models, 
although it is possible (albeit very unlikely) that errors induced by these two assumptions exactly cancel 
out. The new picture we advocate should take into account how gaseous cold flows effectively carry 
larger than previously thought amounts of angular momentum originating from the large-scale motion of 
the cosmic web down to the central region of virialised structures. The fundamental role played by gas 
cooling at low redshift which, for galaxy size halos, preserves this transport of angular momentum in 
spite of the presence of a pervasive shock-heated corona which entirely fills the virialised halo, should be 
described. Finally, since gas accretion at the virial radius is not perfectly aligned with the axis of rotation 
of the galaxy, some angular momentum will inevitably cancel out in the vicinity of  the disc. Careful 
numerical investigation of how angular momentum is transported and/or cancelled in the central region of the 
halo should help shed light into this fundamental aspect of disc formation theory.

\section*{Acknowledgements}
We are grateful to Romain Teyssier for insightful discussions at various stages of this work, 
and Amy McQuillan for a careful reading of an earlier draft of the manuscript. 
We thank Frank van den Bosch, Sadegh Khochfar, Leila Powell, Andrew Benson, Volker Springel, Brad Gibson, and 
Joel Primack  for valuable comments.
The \nut\ simulations, Cosmo25 and Cosmo50 were run on the DiRAC facility jointly funded by STFC, the Large Facilities Capital Fund of BIS and the University of Oxford. Some of the simulations in the \nut\ suite were also run on Jade at the CINES on resources allocated to project number GEN2191. The \mn\ simulation was run at the Barcelona Supercomputing Centre. TK acknowledges support from a Clarendon DPhil studentship. JD and AS's research is supported by Adrian Beecroft, the Oxford Martin School and the STFC. CP thanks Merton College, Oxford for a visiting fellowship allowing the completion of some of this work. This research is part of the Horizon and Horizon-UK project and was also supported by an appointment to the NASA Postdoctoral Program at NASA's Goddard Space Flight Center, administered by Oak Ridge Associated Universities through a contract with NASA. YD is supported by an STFC Postdoctoral Fellowship. We also acknowledge support from the Franco-Korean PHC STAR programme.
\small
\bibliographystyle{mn2e}
\bibliography{refs}

\section*{Appendix: mass accretion histories}

\begin{figure}
  \centering
      \includegraphics[width=8cm]{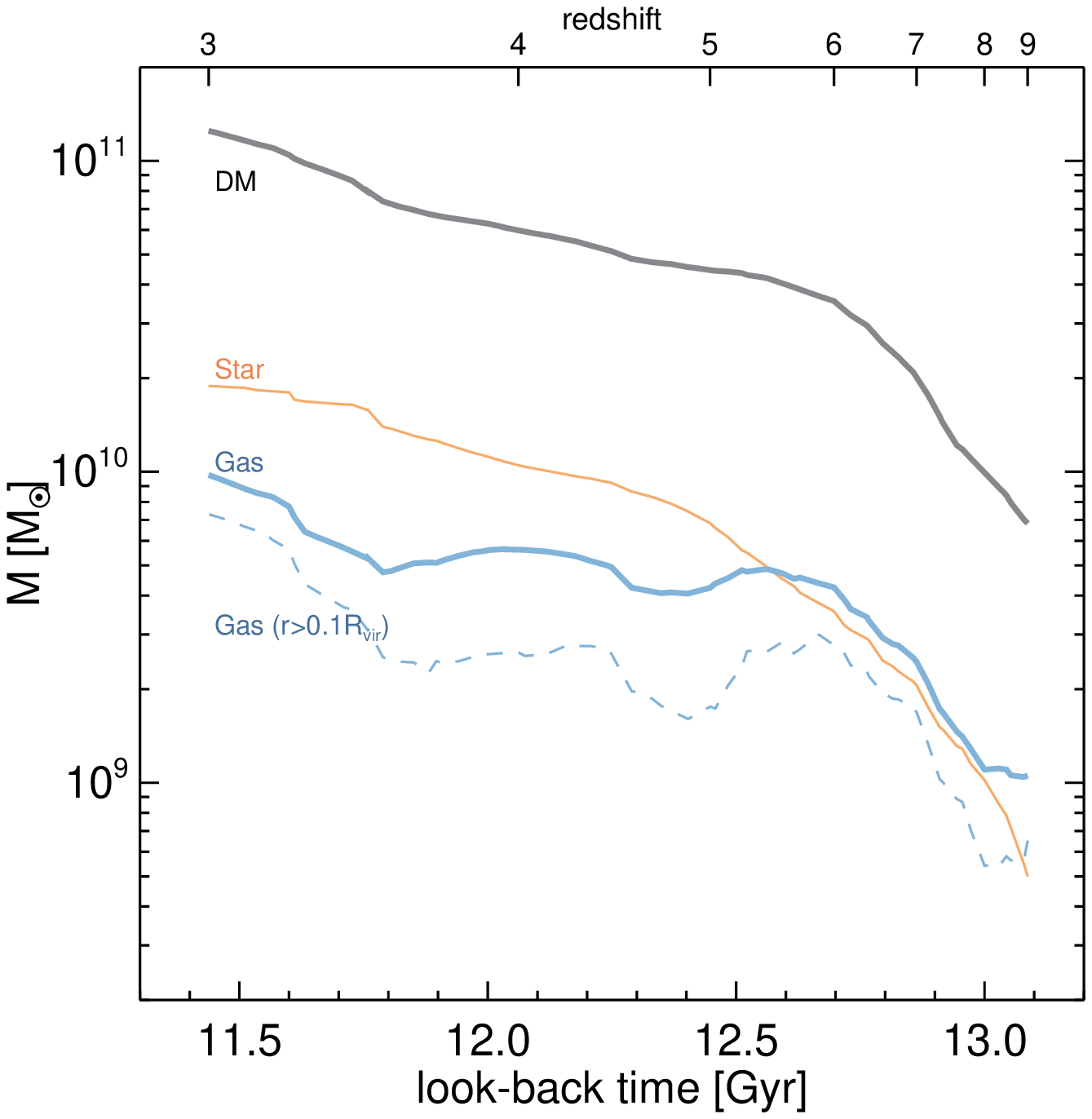} 
     \includegraphics[width=8cm]{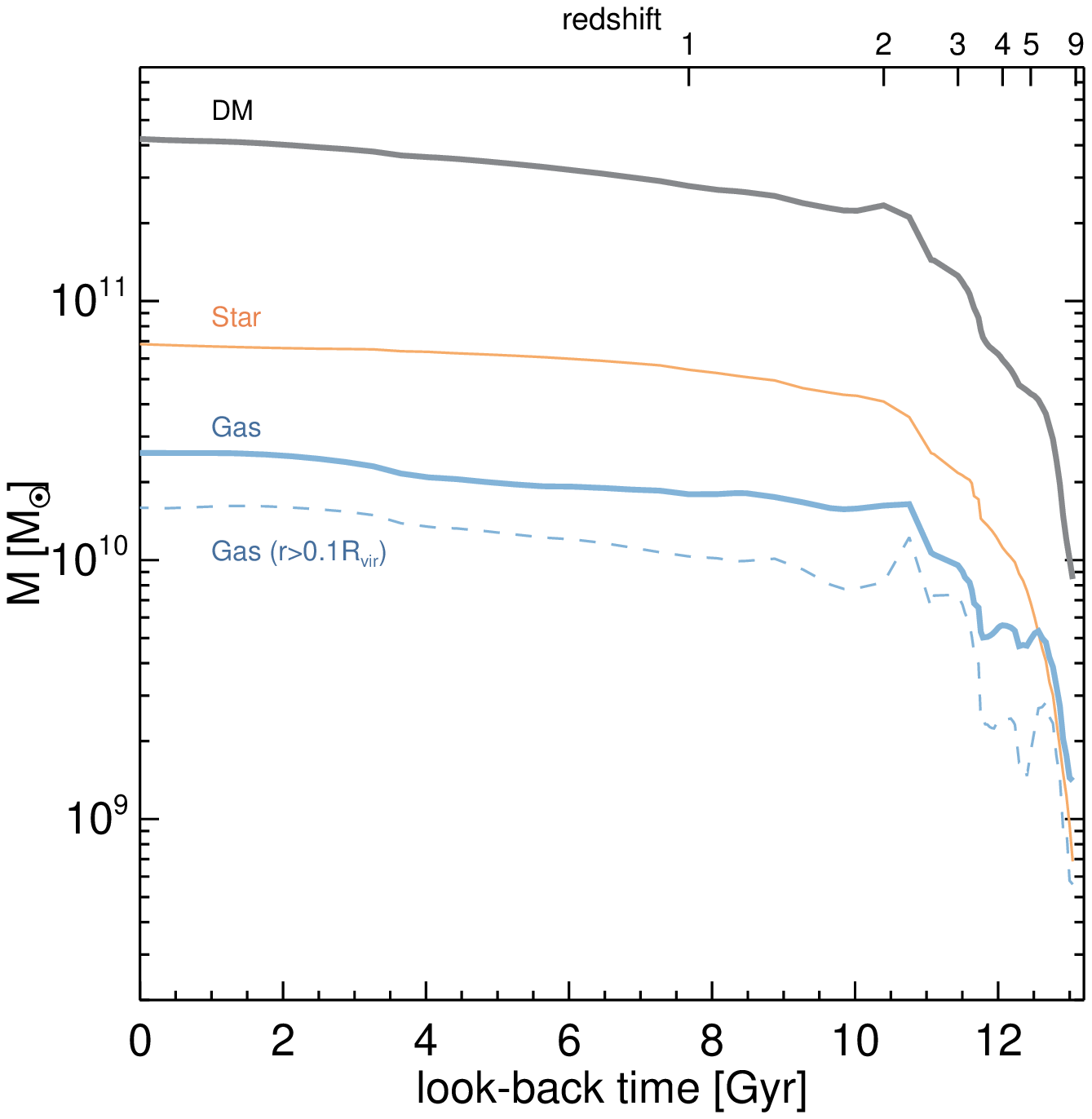} 
 \caption{Mass assembly histories of the various components in the NutFB run (top) and the 
 NutCO run (bottom). Stars, dark matter, and gas in satellite galaxies are included in the measurement.
 Note that the amount of gas mass in the central region ($M_{\rm gas}-M_{\rm gas}(r>0.1\rvir)$) is 
 comparable (a factor 2-3 lower) to the gas mass in the outer region at all times. }
\label{fig:budget}
\end{figure}

In the main body of the text, we show the time evolution of specific angular momentum for the various 
components of a Milky Way-like DM halo (Figs. \ref{fig:jtot} and \ref{fig:jtot_Co}). In particular, in 
Section 3.2, we argue that the difference in angular momentum between the central gas and the gas in the 
outer region of the halo arises because the mass of the central gas, which is composed of low angular 
momentum gas accreted at earlier time, is not negligible compared to the mass of accreting gas, and 
therefore reflects the history of gas accretion. To substantiate this claim, we include the mass assembly 
history of the various components in Fig.~\ref{fig:budget}. As can be seen,  the amount of gas in the 
central region ($M_{\rm gas}-M_{\rm gas}(r>0.1\rvir)$) is comparable to the gas mass in the outer 
region ($r > 0.1\rvir$) at all time. Thus, even if the newly accreted gas carries a larger amount of specific 
angular momentum,  one expects that there will be a non-negligible time-delay for the entire central 
region to be spun at the same level. Finally, note that by multiplying Fig.~\ref{fig:budget} with  Figs. 
\ref{fig:jtot} and \ref{fig:jtot_Co}, one can recover the total amount of angular momentum of each 
component.

\end{document}